\documentclass[acmsmall, manuscript, screen]{acmart}

\AtBeginDocument{%
  }

\setcopyright{acmlicensed}
\copyrightyear{}
\acmYear{}
\acmDOI{XXXXXXX.XXXXXXX}

\usepackage{amsmath}
\usepackage [autostyle, english = american]{csquotes}
\usepackage{extarrows}  
\usepackage{mdframed}
\usepackage{ifthen}
\usepackage{listings}
\usepackage{mathtools}
\usepackage[frozencache=true,cachedir=minted-cache]{minted}
\usepackage{newunicodechar}
\usepackage{phoenician}
\usepackage{textgreek}
\usepackage{semantic}
\usepackage{xspace}

\newunicodechar{⇒}{\ensuremath{\Rightarrow}}
\newunicodechar{→}{\ensuremath{\rightarrow}}
\newunicodechar{λ}{\ifmmode\lambda\else\textlambda\fi}
\newunicodechar{Θ}{\ifmmode\Theta\else\textTheta\fi}
\newunicodechar{Γ}{\ifmmode\Gamma\else\textGamma\fi}
\newunicodechar{μ}{\ifmmode\mu\else\textmu\fi}
\newunicodechar{η}{\ifmmode\eta\else\texteta\fi}
\newunicodechar{Π}{\ifmmode\Pi\else\textPi\fi}
\newunicodechar{𐤄}{\ifmmode{\rotatebox[origin=c]{15}{\textphnc{e}}\hspace{-1pt}}\else{\textphnc{e}}\fi}
\newunicodechar{⊥}{\ensuremath{\bot}}
\newunicodechar{×}{\ensuremath{\times}}
\newunicodechar{⊳}{\ensuremath{\rhd}}
\newunicodechar{∩}{\ensuremath{\cap}}
\newunicodechar{∪}{\ensuremath{\cup}}
\newunicodechar{≠}{\ensuremath{\neq}}
\newunicodechar{∅}{\ensuremath{\varnothing}}
\newunicodechar{⊢}{\ensuremath{\vdash}}
\newunicodechar{△}{\ensuremath{\bigtriangleup}}
\newunicodechar{≡}{\equiv}
\newunicodechar{≢}{\not\equiv}
\newunicodechar{≤}{\ensuremath\leq}
\newunicodechar{⊕}{\oplus}
\newunicodechar{⊖}{\ominus}
\newunicodechar{⟦}{\lBrack}
\newunicodechar{⟧}{\rBrack}
\newunicodechar{∖}{\setminus}

\MakeOuterQuote{"}

\newcommand{\BNF}{\quad\operatorname{::=}\quad}
\newcommand{\BNFOR}{\quad\operatorname{\big{|}}\quad}
\newcommand{\CASE}[6]{\langword{case}_{#1}#2\langword{of}\INL #3 ⇒ #4 ; \INR #5 ⇒ #6}
\newcommand{\CASEm}[6]{\begin{aligned}[t]\langword{case}_{#1}#2\langword{of}&\INL #3 ⇒ #4 ;\\& \INR #5 ⇒ #6\end{aligned}}
\newcommand{\COMM}[2]{\langword{com}_{#1;#2}}
\newcommand{\DEF}{{\quad\operatorname{\triangleq}\quad}}
\newcommand{\DEFCASE}{\quad\operatorname{\xRightarrow{△}}\quad}
\newcommand{\DOT}{\langword{.}}
\newcommand{\eg}{\textit{e.g.}\xspace}
\newcommand{\FLR}[1]{\left\lfloor{#1}\right\rfloor}
\newcommand{\FST}[1]{\langword{fst}_{#1}}
\newcommand{\FuncLang}[1][small]{$𐤄_{λ\mathrm{#1}}$\xspace}
\newcommand{\ie}{\textit{i.e.}\xspace}
\newcommand{\INL}{\langword{Inl}}
\newcommand{\inlinecode}[2][bash]{\mintinline{#1}{#2}}
\newcommand{\INR}{\langword{Inr}}
\newcommand{\langword}[1]{\operatorname{\mathsf{#1}}}
\newcommand{\LOOKUP}[2]{\langword{lookup}^{#1}_{#2}}
\newcommand{\mask}{\ifmmode{\operatorname{⊳}}\else{⊳}\fi\xspace}
\newcommand{\myference}[3]{\inference[\textsc{#1}]{#2}{#3}}
\newcommand{\netstep}[2]{\xlongrightarrow{ #1 \ifthenelse{\equal{#1}{}}{}{:} #2 }}
\newcommand{\nonempty}[1]{{#1^{+}}}
\newcommand{\noop}[2]{\mathtt{noop}^{\mask #1}\!\!(#2)}
\newcommand{\PAIR}{\langword{Pair}}
\newcommand{\prcstep}[2]{\xlongrightarrow{ ⊕#1 ; ⊖#2 }}
\newcommand{\RECV}[1]{\langword{recv}_{#1}}
\newcommand{\roles}[1]{\mathtt{roles}(#1)}
\newcommand{\SEND}[1]{\langword{send}_{#1}}
\newcommand{\set}[1]{\left\{#1\right\}}
\newcommand{\SND}[1]{\langword{snd}_{#1}}
\newcommand{\step}{\operatorname{\longrightarrow}}
\newcommand{\stepname}[1]{$\mathfrak{#1}$}
\newcommand{\vdbl}{\\[8pt]}

\newtheorem{theorem}{Theorem}
\newtheorem{lemma}{Lemma}
\newtheorem{corollary}{Corollary}

\begin{document}

\title[We Know I Know You Know]{We Know I Know You Know; Choreographic Programming With Multicast and Multiply Located Values}

\author{Mako Bates}
\email{mako.bates@uvm.edu}
\orcid{0009-0001-9933-1728}
\affiliation{%
 \institution{University of Vermont}
 \city{Burlington}
 \state{Vermont}
 \country{US}}
\author{Joseph P. Near}
\email{jnear@uvm.edu}
\orcid{0000-0002-3203-3742}
\affiliation{%
 \institution{University of Vermont}
 \city{Burlington}
 \state{Vermont}
 \country{US}}

\begin{abstract}
    Concurrent distributed systems are notoriously difficult to construct and reason about.
    \emph{Choreographic programming} is a recent paradigm that describes a distributed system
    in a single \emph{global} program called a choreography.
    Choreographies simplify reasoning about distributed systems and can ensure deadlock freedom by static analysis.
    In previous choreographic programming languages, each value is \emph{located at} a single party,
    and the programmer is expected to insert special untyped "select" operations
    to ensure that all parties follow the same communication pattern.

    We present \FuncLang, a new choreographic programming language with \emph{multiply located values}.
    \FuncLang allows multicasting to a set of parties, and the resulting value will be located at all of them.
    This approach enables a simple and elegant alternative to "select":
    \FuncLang requires that the guard for a conditional be located at \emph{all} of the relevant parties.
    In \FuncLang, checking that a choreography is well-typed suffices to show that it is deadlock-free.
    We  present several case studies that demonstrate the use of multiply-located values to concisely encode tricky communication patterns
    described in previous work without the use of "select" or redundant communication.

\end{abstract}

\ccsdesc[100]{Theory of computation~Lambda calculus}
\ccsdesc[500]{Theory of computation~Distributed computing models}
\ccsdesc[500]{Computing methodologies~Distributed programming languages}

\keywords{Choreographies, Type Systems, Concurrency, Distributed Systems, Multicast, Broadcast}


\maketitle

\section{Introduction}\label{introduction}

Concurrent distributed systems are notoriously difficult to construct and reason about,
and checking properties like deadlock freedom is particularly challenging.
\emph{Choreographic programming}~\cite{montesi-thesis} is a recent paradigm
that describes a distributed system in a single \emph{global} program called a choreography.
A choreography describes the behavior and communications of all parties
in a single control-flow without the mode-switching characteristic of multi-tier programming.
By making the order and structure of communications explicit, choreographies are deadlock-free by construction.
A process called \emph{endpoint projection} (EPP) compiles the choreography into separate programs for each party
(or participant, process, machine, \textit{etc}) to run;
EPP preserves deadlock freedom and other properties of the original choreography.

One challenge of designing choreographic programming languages is \emph{Knowledge of Choice} (KoC).
In choreographies with conditionals, a KoC strategy ensures that all parties whose behavior depends on the conditional know which branch to take.
Previous choreographic languages require that each value be ``located at'' (\ie known to) a single party,
so only the party where the relevant values are located will know which branch of a conditional to take.
Most choreographic languages provide a special \inlinecode{select} operator
which allows a party with KoC to inform other parties
via additional communication~\cite{choral, chor-lambda-2, chor-lambda, hirsch2021pirouette, montesi-thesis}.
These languages require the programmer to explicitly populate programs with \inlinecode{select} to ensure KoC,
and they rely on their EPP implementations to check if a choreography is well-formed.
An alternative, used by HasChor~\cite{haschor} and ChoRuS~\cite{chorus},
is to communicate the guard of every conditional to all relevant parties---reducing programmer burden compared to \inlinecode{select},
but adding potentially redundant communication to the system.

We present \FuncLang,\footnote{\FuncLang is pronounced "hee lambda small." See Appendix~\ref{sec:name} for more about the name.}
a new choreographic programming language with \emph{multiply-located values}
that avoids redundant communication without the need for a special \inlinecode{select} operator.
In contrast to previous languages, \FuncLang allows each value to be located at a \emph{set} of parties simultaneously.
This new capability enables a simple and elegant solution to KoC---in \FuncLang,
a conditional is well-typed only when the values required to determine KoC are located at \emph{all} relevant parties.
All well-typed \FuncLang choreographies are well-formed.

We present the \FuncLang language and its type system, define endpoint projection for the language,
and prove that it provides deadlock-freedom and that its centralized semantics are correct with respect to EPP.
In several case studies, we show how multiply-located values can be used to concisely encode tricky communication patterns
described in previous work without the use of \inlinecode{select} or redundant communication.

\paragraph{Contributions.}
In summary, we make the following contributions:
\begin{itemize}
\item We introduce \emph{multiply-located values} in choreographic programming.
\item We present \FuncLang, a choreographic programming language that uses multiply-located values
    to ensure KoC without the need for \inlinecode{select}.
\item We define endpoint projection for \FuncLang and prove that it satisfies deadlock-freedom without the traditional partial function "merge".
\item We present several case studies demonstrating the benefit of multiply-located values over previous work.
\end{itemize}

\section{Background}\label{background}

Choreographic programming~\cite{montesi-thesis} is a paradigm that expresses a concurrent distributed system
as a \emph{single} global program describing the behavior and interactions of all parties.
The global view of the distributed system enables easier reasoning about the system's behavior---for example, choreography languages can ensure deadlock-freedom---and also simplify the modular development of complicated interactions between parties.

\subsection{Choreographic Programming}

As a simple example, consider the protocol in Figure~\ref{fig:simple-choreography}, in which a seller wishes to sell a book to a buyer.
The buyer sends the title of the book they want to buy to the seller, and the seller responds with the book's price.
The buyer checks the price against their budget; if they can afford the book, then the seller responds with a date by which it can be delivered.

\begin{figure}[tbhp]
    \begin{mdframed}
\begin{minted}[xleftmargin=10pt,linenos]{bash}
seller.title <- buyer.title                          # buyer sends title to seller
buyer.price <- seller.getPrice(seller.title)         # seller sends price to buyer
if buyer.budget > buyer.price
  then: buyer.date <- seller.getDate(seller.title)   # sellers sends date to buyer
\end{minted}
    \caption{A simple choreography between a buyer and seller.}
    \label{fig:simple-choreography}
    \Description[A four-line pseudo-code choreography of the classic "bookseller protocol".]
                {A four-line pseudo-code choreography of the classic "bookseller protocol".
                There are two parties, a seller and a buyer. There is one "if" branch, based on data the buyer owns.}
    \end{mdframed}
\end{figure}

This simple example demonstrates the main features of choreographic programming. It mixes communication (using the \inlinecode{<-} operator to send values) with computation (e.g. the \inlinecode{getPrice} function to compute the price of a book) in a single global program. In a choreographic program, each value has a \emph{location} indicating which party stores the value (e.g. \inlinecode{seller.title} is located at the seller, while \inlinecode{buyer.title} is located at the buyer).

\subsection{Endpoint Projection}\label{sec:background-epp}

Executing a choreography requires compiling it into separate programs for each of the parties to run---a process called \emph{endpoint projection} (EPP)~\cite{montesi-thesis}.
EPP "projects" a choreography to an "endpoint" (process, machine, location, \textit{etc})
in a sense analogous to geometric projection of a high-dimension object to its lower-dimensional shadow.
For example, EPP can transform the program in Figure~\ref{fig:simple-choreography} into two separate programs---one for the buyer and one for the seller---as shown in Figure~\ref{fig:epp}.

\begin{figure}[tbhp]
  \begin{mdframed}
    \centering
  \begin{tabular}{l l}
    \begin{minipage}{.45\textwidth}
    \begin{mdframed}[linecolor=lightgray]
\begin{minted}[xleftmargin=10pt,linenos]{bash}
send(title, seller)
price = recv(seller)
if budget > price
  then: date = recv(seller)
\end{minted}
\end{mdframed}
\centering \textbf{Buyer}
\end{minipage}
    &
    \begin{minipage}{.45\textwidth}
    \begin{mdframed}[linecolor=lightgray]
\begin{minted}[xleftmargin=10pt,linenos]{bash}
title = recv(buyer)
send(getPrice(title), buyer)
if ????
  then: send(getDate(title), buyer)
\end{minted}
\end{mdframed}
\centering \textbf{Seller}
\end{minipage}
  \end{tabular}
    \caption{Endpoint projection of the example from Figure~\ref{fig:simple-choreography}.}
    \label{fig:epp}
    \Description[Two different local-process programs, one for the Seller and one for the Buyer.]
                {Two different local-process programs, one for the Seller and one for the Buyer.
                They're both four-lines long, and have similar structure.
                If run together, they would implement the original choreography,
                except that it's unclear what the guard should be to "if" in the Seller's program.}
  \end{mdframed}
\end{figure}

Endpoint projection translates each statement from the choreography in Figure~\ref{fig:simple-choreography}
into a corresponding statement for the specified party to run.
Each communication in the original choreography becomes a call to \inlinecode{send} for the original sender
and a \inlinecode{recv} for the original receiver.
These functions can be implemented with traditional network primitives like blocking sockets;
since the original choreography exactly specifies the sequence of communications, the projected program will not contain deadlocks.

Choreographies with conditionals---like the Bookseller example---introduce a challenge for endpoint projection:
\emph{some parties might not know which branch to take!}
In this example, the final communication occurs only if the price of the book is within the buyer's budget,
but the budget value is located at the buyer and not known to the seller.

\subsection{Knowledge of Choice}

To address this challenge, all choreographic programming languages include a strategy for \emph{Knowledge of Choice} (KoC), which ensures that relevant parties have enough information to determine the communication structure of the program.

The most common KoC strategy is to syntactically ensure that
each branching operation is controlled by a single party,
and that they communicate their choice to other relevant parties using
a designated \inlinecode{select} operation~\cite{choral, chor-lambda-2, chor-lambda, hirsch2021pirouette}.
In these languages, the programmer is expected to ensure KoC explicitly.
EPP is will fail for choreographies without correct KoC management;
this guards against implementation mistakes in advance of runtime,
but type systems used in these languages do not check if EPP is defined.

Figure~\ref{fig:epp2} adapts the example from Figure~\ref{fig:simple-choreography} to the syntax of
Chor$\lambda$~\cite{chor-lambda} and uses \inlinecode{select} for KoC.
As in Chor$\lambda$, \inlinecode{com[p][q] x} denotes that \inlinecode{p} communicates \inlinecode{x} to \inlinecode{q},
and \inlinecode{select[p][q] l} denotes that \inlinecode{p} communicates the choice \inlinecode{l} to \inlinecode{q}.
In our example, the buyer would use \inlinecode{select} to inform the seller of the conditional's result by communicating a single boolean flag.
In the projected programs, the buyer's act of sending the flag appears very similar to the choreographic representation,
and the seller's reception of the flag is denoted using \inlinecode{offer}, as shown in Figure~\ref{fig:epp2}.
(The \inlinecode{offer} and \inlinecode{choose} syntax comes from multi-party-session-types.)

HasChor~\cite{haschor} solves the KoC problem by broadcasting the chosen branch of each conditional to all parties. This approach reduces programmer burden but can result in unneeded additional communication.

\begin{figure}
  \begin{mdframed}
    \centering
    \begin{mdframed}[linecolor=lightgray]
\begin{minted}[xleftmargin=10pt,linenos]{bash}
let title = com[buyer][seller] buyer_title;      # buyer sends title to seller
let price = com[seller][buyer] getPrice(title);  # seller sends price to buyer
case (budget > price) of
  True  => select[buyer][seller] ok;             # buyer communicates choice
           com[seller][buyer] getDate(title);    # sellers sends date to buyer
  False => select[buyer][seller] ko;             # buyer communicates choice
\end{minted}
\end{mdframed}

\bigskip
$\Downarrow$
\bigskip

  \begin{tabular}{l l}
    \begin{minipage}{.48\textwidth}
    \begin{mdframed}[linecolor=lightgray]
\begin{minted}[xleftmargin=10pt,linenos]{bash}
let _ = send[seller] buyer_title;
let price = recv[seller];
case (budget > price) of
  True  => let _ = choose[seller] ok;
           recv[seller];
  False => choose[seller] ko;
\end{minted}
\end{mdframed}
\centering \textbf{Buyer}
\end{minipage}
    &
    \begin{minipage}{.48\textwidth}
    \begin{mdframed}[linecolor=lightgray]
\begin{minted}[xleftmargin=10pt,linenos]{bash}
let title = recv[buyer];
let _ = send[buyer] getPrice(title);
offer[buyer] {
  ok => send[buyer] getDate(title);
  ko => ()
}
\end{minted}
\end{mdframed}
\centering \textbf{Seller}
\end{minipage}
  \end{tabular}
    \caption{A simple choreography between a buyer and seller, made projectable using \inlinecode{select} (top), and its projection (bottom).
      This example adapts Figure~\ref{fig:simple-choreography} to the syntax of Chor$\lambda$;
      \inlinecode{<-} becomes \inlinecode{com}, \inlinecode{if} becomes \inlinecode{case},
      and added calls to \inlinecode{select} for KoC project as \inlinecode{offer} and \inlinecode{choose}.}
    \label{fig:epp2}
    \Description[New versions of the Bookseller Protocol and its two endpoint projections, in Chor-Lambda syntax using "select".]
                {New versions of the Bookseller Protocol and its two endpoint projections.
                These versions are in Chor-Lambda syntax and use "select" to fix the Knowledge of Choice problem.
                During EPP, "select" transforms into the "offer" and "choose" syntax from multi-party-session types.}
  \end{mdframed}
\end{figure}

\section{Choreographies Without "Select"}\label{sec:contribution}

Our work presents an alternative approach for KoC that eliminates the need for \inlinecode{select} and does not require redundant communication. Our approach is based on two insights:
\begin{itemize}
  \item If only parties who can evaluate the guard-value of a conditional
      participate in its branches, then no additional communication is needed
        for KoC.
    \item If \inlinecode{p} sends value \inlinecode{X} to \inlinecode{q},
        then \emph{both \inlinecode{p} and \inlinecode{q} know \inlinecode{X}}.
\end{itemize}

We leverage these insights in a new choreographic language called \FuncLang. In \FuncLang, each value is \emph{multiply located} and the communication operator (\inlinecode{com}) is implemented as a multicast operator. To ensure KoC, \FuncLang's type system ensures that a conditional's guard is located at \emph{all} relevant parties.
Specifically:
\begin{enumerate}
  \item Data is multiply-located.
      Rather than having a single owner, data values (and functions) are owned
        (and known to) non-empty sets of parties, \eg $()@\{p, q\}$ is unit located at $p$ and $q$.
    \item For a case expression such as
        $\;\CASE{\{p, q\}}{V}{x_l}{M_l}{x_r}{M_r}\;$
        to type-check, $V$ must be known to both $p$ and $q$,
        and only $p$ and $q$ may participate in the branches $M_l$ and $M_r$.
    \item The $\COMM{s}{\nonempty{r}}$ built-in function is
        a multicast operator;
        it returns a multiply-located value at all parties in the set $\nonempty{r}$
        (which may include $s$).
\end{enumerate}

\begin{figure}
  \begin{mdframed}
    \centering
    \begin{mdframed}[linecolor=lightgray]
\begin{minted}[xleftmargin=7pt,linenos]{bash}
let title = com[buyer][seller] buyer_title;        # buyer sends title to seller
let price = com[seller][buyer] getPrice(title);    # seller sends price to buyer
case com[buyer][buyer,seller] (budget > price) of  # buyer multicasts choice
  True  => com[seller][buyer] getDate(title);      # seller sends date to buyer
  False => ()
\end{minted}
\end{mdframed}

\bigskip
$\Downarrow$
\bigskip

  \begin{tabular}{l l}
      \!\!\!\!\!\!\!\!
    \begin{minipage}{.53\textwidth}
    \begin{mdframed}[linecolor=lightgray]
\begin{minted}[xleftmargin=7pt,linenos]{bash}
let _ = send[seller] buyer_title;
let price = recv[seller];
case send[buyer,seller] (budget > price) of
  True  => recv[seller];
  False => ()
\end{minted}
\end{mdframed}
\centering \textbf{Buyer}
\end{minipage}
    &
    \begin{minipage}{.47\textwidth}
    \begin{mdframed}[linecolor=lightgray]
\begin{minted}[xleftmargin=7pt,linenos]{bash}
let title = recv[buyer];
let _ = send[buyer] getPrice(title);
case recv[buyer] of
  True  => send[buyer] getDate(title);
  False => ()
\end{minted}
\end{mdframed}
\centering \textbf{Seller}
\end{minipage}
  \end{tabular}
\caption{The buyer and seller example from Figure~\ref{fig:epp2}, written in \FuncLang without \inlinecode{select}. In line 3, the \inlinecode{com} function multicasts the conditional's guard to both parties, ensuring KoC for the conditional. The multicast \inlinecode{com} operator is transformed into a multicast \inlinecode{send} during endpoint projection.}
    \label{fig:epp3}
    \Description[New versions of the Bookseller Protocol and its two endpoint projections in He-Lambda-Small syntax using multicast "com".]
                {New versions of the Bookseller Protocol and its two endpoint projections in He-Lambda-Small syntax.
                 This version uses multicast "com" to solve the Knowledge of Choice problem;
                 the EPP of a multicast "com" is just a multicast "send" or a perfectly normal "receive".}
  \end{mdframed}
\end{figure}

\subsection{Multiply-located values}\label{sec:multiply-located}
Previous choreography languages have featured \emph{located values},
values annotated with (or implicitly assigned to) their owning party such that EPP to the owner results
in the value itself and EPP to any other party results in a special
"missing" value (\eg ⊥).
\emph{Multiply located values} are exactly the same except they are annotated with a non-empty \emph{set} of parties.
In \FuncLang, the EPP of a multiply-located value is the same for all owning parties, and ⊥ for other parties.
Including multiply-located values as first-class syntax in \FuncLang
works well with the multicast-style $\COMM{p}{\nonempty{q}}$ operator.
Prior works have objects with multiple owners as emergent structures in a language
(\eg choreographic processes~\cite{choral}, distributed choice types~\cite{chor-lambda-2}),
but these project to each owner's distinct view of the structure.

Multiply-located values also enable concise expression of programs in which multiple parties compute the same thing in parallel---a common occurrence when communication is more expensive than computation. For example, the expression $5@\set{p,q,r}+3@\set{p,q,r}$ represents an addition performed by all three parties in parallel.

\section{The \FuncLang Language}\label{sec:language}

This section presents the \FuncLang language.
Its syntax and semantics are loosely based on
Chorλ~\cite{chor-lambda},
but to simplify presentation we omit recursion and polymorphism.
In Sections~\ref{sec:syntax} through~\ref{sec:semantics}
we describe the syntax, type system, and centralized semantics of \FuncLang.
As in other choreographic languages, the centralized semantics describe the intended meaning of choreographies
and can be used to reason about their behavior.
Sections~\ref{sec:local-lang} through~\ref{sec:networks} describe the semantics of distributed processes and define endpoint projection for \FuncLang.
In Section~\ref{sec:deadlock-freedom}, we prove that the behavior of a projected choreography matches that of the original choreography under the centralized semantics, and that \FuncLang ensures deadlock-freedom.

\subsection{Syntax}\label{sec:syntax}
The syntax of \FuncLang is in Figure~\ref{fig:syntax}.
Location information sufficient for typing, semantics, and EPP is explicit in
the expression forms.
We distinguish between "pairs"
($\PAIR V_1 V_2$, of type $(d_1 × d_2)@\nonempty{p}$)
and "tuples"
($(V_1, V_2)$, of type $(T_1, T_2)$)
so that we can have a distinguishable concept of "data" as "stuff that can be sent";
we do not believe this to have any theoretic significance.
Throughout this text we assume bound variables are unique;
any implementation of \FuncLang should use normal techniques
to uniquify variables before evaluation or EPP.

The superscript-marked identifier $\nonempty{p}$ is a single token representing a set of parties;
an unmarked $p$ is a completely distinct token representing a single party.
Note the use of a superscript "$+$" to denote sets of parties
instead of a hat or boldface;
this denotes that these lists may never be empty.\footnote{
Later, we'll use an "$\ast$" to denote a possibly-empty set or list,
and (in the appendices) a "$?$" to denote "zero or one".
}
The type and semantic rules will enforce this invariant as needed.
When a set of parties should be understood as "context"
rather than "attribute" (\eg in the typing rules),
we write Θ rather than $\nonempty{p}$;
this is entirely to clarify intent and the distinction has no formal meaning.

\begin{figure}[tbhp]
    \begin{mdframed}
\begin{align*}
M  \BNF   &  V                       && \text{Values.}          \\
   \BNFOR &  M M                     && \text{Function application.}          \\
   \BNFOR &  \CASE{\nonempty{p}}{M}{x}{M}{x}{M}    \quad&& \text{Branching on a disjoint-sum value.}          \\
                                            \\
V  \BNF   &  x                       && \text{Variables.}          \\
   \BNFOR &  (λ x:T \DOT M)@\nonempty{p}            && \text{Function literals annotated with participants.}          \\
   \BNFOR &  ()@\nonempty{p}                      && \text{Multiply-located unit.}          \\
   \BNFOR &  \INL V                  && \text{Injection to a disjoint-sum.}           \\
   \BNFOR &  \INR V                  && \text{}           \\
    \BNFOR &  \PAIR V V               && \text{Construction of data pairs (products).}           \\
   \BNFOR &  (V, \dots, V)           && \text{Construction of heterogeneous tuples.}           \\
   \BNFOR &  \FST{\nonempty{p}}      && \text{Projection of data pairs.}           \\
   \BNFOR &  \SND{\nonempty{p}}      && \text{}           \\
   \BNFOR &  \LOOKUP{n}{\nonempty{p}}   && \text{Projection of tuples.}           \\
   \BNFOR &  \COMM{p}{\nonempty{p}}     && \text{Send to one or more recipients.}            \\
                                            \\
d  \BNF   &  ()         && \text{We provide a simple algebra of "data" types,}   \\
   \BNFOR &  d + d                   && \text{which can encode booleans or other finite types}           \\
   \BNFOR &  d × d                   && \text{and could be extended with natural numbers if desired.}   \\
                                            \\
T  \BNF   &  d@\nonempty{p}          && \text{A complete multiply-located data type.}             \\
    \BNFOR &  (T → T)@\nonempty{p}          && \text{Functions are located at their participants.}             \\
   \BNFOR &  (T, \dots, T)           && \text{A fixed-length heterogeneous tuple.}  \\
\end{align*}
    \caption{The complete syntax of the \FuncLang language.}
    \label{fig:syntax}
    \Description[A BNF syntax for a choreographic lambda calculus.]
                {A BNF syntax for a choreographic lambda calculus.
                There are three expression-forms M,
                including the V form for values,
                of which there are eleven sub-forms.
                There are also forms for types.
                Both types and expressions have party-annotations.}
    \end{mdframed}
\end{figure}

\subsection{The Mask Operator}\label{sec:masking}
Here we introduce the \mask operator,
the purpose of which is to allow Theorem~\ref{theorem:preservation} to hold
without adding sub-typing or polymorphism to \FuncLang.
\mask is a partial function defined in Figure~\ref{fig:masking};
the left-hand argument is either a type (in which case it returns a type)
or a value (in which case it returns a value).
The effect of \mask is very similar to EPP,
except that it projects to a set of parties instead of just one,
and instead of introducing a ⊥ symbol it is simply undefined in some cases.
Because it is used during type-checking, errors related to it are caught at that time.

Consider an expression using a "masking identity" function:
$(λ x:()@\set{p} \DOT x)@\set{p} ()@\set{p,q}$,
where the lambda is an identity function \emph{application of which}
turns a multiply-located unit value into one located at just $p$.
Clearly, the lambda should type as $(()@\set{p} → ()@\set{p})@\set{p}$;
and so the whole application expression should type as $()@\set{p}$.
Masking in the typing rules lets this work as expected,
and similar masking in the semantic rules ensures type preservation.

\begin{figure}[tbhp]
    \begin{mdframed}
\begin{gather*}
\myference{MTData}
          {\nonempty{p} ∩ Θ ≠ ∅}
          {d@\nonempty{p} \mask Θ \DEF d@(\nonempty{p} ∩ Θ)}
          \quad
\myference{MTFunction}
          {\nonempty{p} \subseteq Θ}
          {(T → T')@\nonempty{p} \mask Θ \DEF (T → T')@\nonempty{p}}
          \vdbl
\myference{MTVector}
          {T_1' = T_1 \mask Θ, \quad \dots \quad T_n' = T_n \mask Θ}
          {(T_1, \dots, T_n) \mask Θ \DEF (T_1', \dots, T_n')}
          \vdbl
\myference{MVLambda}
          {\nonempty{p} \subseteq Θ}
          {(λ x:T \DOT M)@\nonempty{p} \mask Θ \DEF (λ x:T \DOT M)@\nonempty{p}}
          \quad
\myference{MVUnit}
          {\nonempty{p} ∩ Θ ≠ ∅}
          {()@\nonempty{p} \mask Θ \DEF ()@(\nonempty{p} ∩ Θ)}
          \vdbl
\myference{MVInL}
          {V' = V \mask Θ}
          {\INL V \mask Θ \DEF \INL V'}
          \quad
\myference{MVInR}
          {\dots}
          {\dots}
          \quad
\myference{MVProj1}
          {\nonempty{p} \subseteq Θ}
          {\FST{\nonempty{p}} \mask Θ \DEF \FST{\nonempty{p}}}
          \quad
\myference{MVProj2}
          {\dots}
          {\dots}
          \vdbl
\myference{MVPair}
          {V_1' = V_1 \mask Θ \quad V_2' = V_2 \mask Θ}
          {\PAIR V_1 V_2 \mask Θ \DEF \PAIR V_1' V_2'}
          \quad
\myference{MVVector}
          {V_1' = V_1 \mask Θ \quad \dots \quad V_n' = V_n \mask Θ}
          {(V_1, \dots, V_n) \mask Θ \DEF (V_1', \dots, V_n')}
          \vdbl
\myference{MVProjN}
          {\nonempty{p} \subseteq Θ}
          {\LOOKUP{n}{\nonempty{p}} \mask Θ \DEF \LOOKUP{n}{\nonempty{p}}}
          \quad
\myference{MVCom}
          {s \in Θ \quad \nonempty{r} \subseteq Θ}
          {\COMM{s}{\nonempty{r}} \mask Θ \DEF \COMM{s}{\nonempty{r}}}
          \quad
\myference{MVVar}
          {}
          {x \mask Θ \DEF x}
\end{gather*}
    \caption{Definition of the \mask operator.}
    \label{fig:masking}
    \Description[Inference rules defining a partial function "mask".]
                {Inference rules defining a partial function "mask"
                denoted by a rightward triangle.
                The right-hand argument is a non-empty set of parties,
                and the left-hand argument is either a type or a value
                in he-lambda-small.
                It's defined as the left-hand argument re-located
                to the right-hand argument, provided the new locations
                are a subset of the original locations and that the new value
                is still semantically useable.}
    \end{mdframed}
\end{figure}

\subsection{Typing Rules}\label{sec:typing}
The typing rules for \FuncLang are in Figure~\ref{fig:typing}.
A judgment $Θ;Γ ⊢ M : T$ says that $M$ has type $T$ in the context
of a non-empty set of participating parties Θ
and a (possibly empty) list of variable bindings $Γ=(x_1:T_1), \dots (x_n:T_n)$.
In \textsc{TLambda} and \textsc{TProjN} we write preconditions
$\noop{\nonempty{p}}{T}$ meaning $T = T \mask \nonempty{p}$,
\ie masking to those parties is a "no-op".
We are consistently assuming that bound variables are unique;
the freshness of $x$, $x_l$, and $x_r$ in \textsc{TLambda} and \textsc{TCase}
may be considered as extra implicit preconditions.

\begin{figure}[tbhp]
    \begin{mdframed}
\begin{gather*}
\myference{TLambda}
          {\nonempty{p};Γ,(x:T) ⊢ M : T' \quad
           \nonempty{p} \subseteq Θ \quad
           \noop{\nonempty{p}}{T}}
          {Θ;Γ ⊢ (λ x:T \DOT M)@\nonempty{p} : (T → T')@\nonempty{p}}
          \quad
\myference{TVar}
          {x : T \in Γ \quad T' = T \mask Θ}
          {Θ;Γ ⊢ x : T' }
          \vdbl
\myference{TApp}
          {Θ;Γ ⊢ M : (T_a → T_r)@\nonempty{p} \quad
           Θ;Γ ⊢ N : T_a' \quad
           T_a' \mask \nonempty{p} = T_a}
          {Θ;Γ ⊢ M N : T_r}
          \vdbl
\myference{TCase}
          {Θ;Γ ⊢ N : T_N \quad
           (d_l + d_r)@\nonempty{p} = T_N \mask \nonempty{p} \\
           \nonempty{p};Γ,(x_l:d_l@\nonempty{p}) ⊢ M_l : T \quad
           \nonempty{p};Γ,(x_r:d_r@\nonempty{p}) ⊢ M_r : T \quad
           \nonempty{p} \subseteq Θ}
          {Θ;Γ ⊢ \CASE{\nonempty{p}}{N}{x_l}{M_l}{x_r}{M_r} : T}
          \vdbl
\myference{TUnit}
          {\nonempty{p} \subseteq Θ}
          {Θ;Γ ⊢ ()@\nonempty{p} : ()@\nonempty{p}}
          \quad
\myference{TPair}
          {Θ;Γ ⊢ V_1 : d_1@\nonempty{p_1} \quad
           Θ;Γ ⊢ V_2 : d_2@\nonempty{p_2} \quad
           \nonempty{p_1} ∩ \nonempty{p_2} ≠ ∅}
          {Θ;Γ ⊢ \PAIR V_1 V_2 : (d_1 × d_2)@(\nonempty{p_1} ∩ \nonempty{p_2})}
          \vdbl
\myference{TVec}
          {Θ;Γ ⊢ V_1 : T_1 \quad \dots \quad Θ;Γ ⊢ V_n : T_n}
          {Θ;Γ ⊢ (V_1, \dots, V_n) : (T_1, \dots T_n)}
          \quad
\myference{TInl}
          {Θ;Γ ⊢ V : d@\nonempty{p}}
          {Θ;Γ ⊢ \INL V : (d + d')@\nonempty{p}}
          \quad
\myference{TInr}{\dots}{\dots}
          \vdbl
\myference{TProjN}
          {\nonempty{p} \subseteq Θ \quad
           \noop{\nonempty{p}}{(T_1, \dots, T_n)}}
          {Θ;Γ ⊢ \LOOKUP{i}{\nonempty{p}} : ((T_1, \dots, T_i, \dots, T_n) → T_i)@\nonempty{p}}
          \quad
\myference{TProj2}{\dots}{\dots}
          \vdbl
\myference{TProj1}
          {\nonempty{p} \subseteq Θ}
          {Θ;Γ ⊢ \FST{\nonempty{p}} : ((d_1 × d_2)@\nonempty{p} → d_1@\nonempty{p})@\nonempty{p}}
          \quad
\myference{TCom}
          {s \in \nonempty{s} \quad
           \nonempty{s}\cup\nonempty{r} \subseteq Θ}
          {Θ;Γ ⊢ \COMM{s}{\nonempty{r}} : (d@\nonempty{s} → d@\nonempty{r})@(\set{s}\cup\nonempty{r})}
\end{gather*}
    \caption{\FuncLang typing rules.}
    \label{fig:typing}
    \Description[Inference rules for he-lambda-small's type system.]
                {Inference rules for he-lambda-small's type system.
                There are thirteen of them, corresponding to the thirteen
                total expression forms.}
    \end{mdframed}
\end{figure}

Examine \textsc{TCase} as the most involved example.
The actual judgment says that in the context of Θ and Γ,
the case expression types as $T$.
The first two preconditions say that
the guard expression $N$ must type in the parent context
as some type $T_N$, which masks to the explicit party-set $\nonempty{p}$
as a sum-type $(d_l + d_r)@\nonempty{p}$.
The only rule by which it can do that is \textsc{MTData},
so we can deduce that $T_N = (d_l + d_r)@\nonempty{q}$,
where $\nonempty{q}$ is some unspecified superset of $\nonempty{p}$.
The third and forth preconditions say that $M_l$ and $M_r$
must both type as $T$ in the context of $\nonempty{p}$ instead of Θ
and with the respective $x_l$ and $x_r$ bound to the right and left
data types at $\nonempty{p}$.
The final precondition says that $\nonempty{p}$ is a subset of Θ,
\ie everyone who's supposed to be branching is actually present to do so.

The other rules are mostly normal, with similar masking of types and narrowing
of participant sets as needed.
In \textsc{TVar}, the Θ context overrides (masks) the type bindings in Γ.
In isolation, some expressions such as $\INR ()@\set{p}$
or the projection operators
are flexible about their exact types;
additional parameters could give them monomorphic typing,
if that was desirable.

\subsection{Substitution in \FuncLang}\label{sec:substitution}

For \mask to fulfil its purpose during semantic evaluation,
it may need to be applied arbitrarily many times with different party-sets
inside the new expressions, and it may not even be defined for all such
party-sets.
Conceptually, this just recapitulates the masking performed in \textsc{TVar}.
To formalize these subtleties, in Figure~\ref{fig:substitution} we specialize the normal variable-substitution
notation $M[x:=V]$ to perform location-aware substitution.
Theorem~\ref{theorem:substitution} shows that this operation
satisfies the usual concept of substitution.

\begin{theorem}[Substitution]\label{theorem:substitution}
  If $Θ;Γ,(x:T_x) ⊢ M : T$ and $Θ;Γ ⊢ V : T_x$,
  then $Θ;Γ ⊢ M[x := V] : T$.

  See Appendix~\ref{sec:substitution-proof} for the proof.
\end{theorem}

\begin{figure}[tbhp]
    \begin{mdframed}
\begin{align*}
M[x:=V] \DEF \text{by pattern matching on $M$:}& \\
y            \DEFCASE & \begin{cases}
                                        y ≡ x & \DEFCASE  V  \\
                                        y ≢ x & \DEFCASE  y
                                        \end {cases} \\
N_1 N_2     \DEFCASE & N_1[x:=V] N_2[x:=V] \\
(λ y:T \DOT N)@\nonempty{p}  \DEFCASE & \begin{cases}
                                        V \mask \nonempty{p} = V'
                                            & \DEFCASE (λ y:T \DOT N[x:=V'])@\nonempty{p} \\
                                        \text{otherwise} & \DEFCASE M
                                        \end{cases} \\
\CASEm{\nonempty{p}}{N}{x_l}{M_l}{x_r}{M_r} \DEFCASE & \begin{cases}
                                        V \mask \nonempty{p} = V'
                                            & \DEFCASE \CASEm{\nonempty{p}}
                                                            {N[x:=V]}{x_l}{M_l[x:=V']}
                                                            {x_r}{M_r[x:=V']} \\
                                        \text{otherwise}
                                            & \DEFCASE \CASEm{\nonempty{p}}
                                                            {N[x:=V]}{x_l}{M_l}{x_r}{M_r}
                                        \end{cases} \\
\INL V_1    \DEFCASE & \INL V_1[x:=V] \\
\INR V_2    \DEFCASE & \INR V_2[x:=V] \\
\PAIR V_1 V_2  \DEFCASE & \PAIR V_1[x:=V] V_2[x:=V] \\
(V_1, \dots, V_n) \DEFCASE & (V_1[x:=V], \dots, V_n[x:=V]) \\
\begin{rcases}
    ()@\nonempty{p}
    \qquad \FST{\nonempty{p}}
    \qquad \SND{\nonempty{p}} \\
    \qquad \LOOKUP{\nonempty{p}}{i}
    \qquad \COMM{s}{\nonempty{r}}
\end{rcases}\DEFCASE & M
\end{align*}
    \caption{The customised substitution used in \FuncLang's semantics.}
    \label{fig:substitution}
    \Description[A case-wise definition of variable substitution.]
                {A case-wise definition of variable substitution.
                 Most cases are normal; some involve masking
                 and when the masking is undefined they revert to a no-op.}
    \end{mdframed}
\end{figure}

\subsection{Centralized Semantics}\label{sec:semantics}

The semantic stepping rules for evaluating \FuncLang expressions
in the central model (\ie semantic stepping for choreographies \textit{per se},
with all notions of local processes and communication between them left implicit)
are in Figure~\ref{fig:semantics}.
In Sections~\ref{sec:local-lang}, \ref{sec:projection}, and \ref{sec:networks}
we will develop the "ground truth" of the distributed process semantics and show that
the centralized semantics correctly capture distributed behavior.

\FuncLang is equipped with a substitution-based semantics that,
after accounting for the \mask operator and the specialized implementation of
substitution, is quite standard among lambda-calculi.
In particular, we make no effort here to support the out-of-order execution
supported by some choreography languages.
Because the language and corresponding computational model are parsimonious,
no step-annotations are needed for the centralized semantics.

\begin{figure}[tbhp]
    \begin{mdframed}
\begin{gather*}
\myference{AppAbs}
          {V' = V \mask \nonempty{p}}
          {((λ x:T \DOT M)@\nonempty{p}) V \step M[x := V']}
          \quad
\myference{App1}
          {N \step N'}
          {V N \step V N'}
          \quad
\myference{App2}
          {M \step M'}
          {M N \step M' N}
          \vdbl
\myference{Case}
          {N \step N'}
          {\CASE{\nonempty{p}}{N}{x_l}{M_l}{x_r}{M_r}
            \step \CASE{\nonempty{p}}{N'}{x_l}{M_l}{x_r}{M_r}}
          \vdbl
\myference{CaseL}
          {V' = V \mask \nonempty{p}}
          {\CASE{\nonempty{p}}{\INL V}{x_l}{M_l}{x_r}{M_r} \step M_l[x_l := V']}
          \quad
\myference{CaseR}
          {\dots}
          {\dots}
          \vdbl
\myference{Proj1}
          {V' = V_1 \mask \nonempty{p}}
          {\FST{\nonempty{p}} (\PAIR V_1 V_2) \step V'}
          \quad
\myference{Proj2}
          {\dots}
          {\dots}
          \quad
\myference{ProjN}
          {V' = V_i \mask \nonempty{p}}
          {\LOOKUP{i}{\nonempty{p}} (V_1, \dots, V_i, \dots, V_n) \step V'}
          \vdbl
\myference{Com1}
          {()@\nonempty{p} \mask \set{s} = ()@s}
          {\COMM{s}{\nonempty{r}} ()@\nonempty{p} \step ()@\nonempty{r}}
          \quad
\myference{ComPair}
          {\COMM{s}{\nonempty{r}} V_1 \step V_1' \quad \COMM{s}{\nonempty{r}} V_2 \step V_2'}
          {\COMM{s}{\nonempty{r}} (\PAIR V_1 V_2) \step \PAIR V_1' V_2'}
          \vdbl
\myference{ComInl}
          {\COMM{s}{\nonempty{r}} V \step V'}
          {\COMM{s}{\nonempty{r}} (\INL V) \step \INL V'}
          \quad
\myference{ComInr}
          {\dots}
          {\dots}
\end{gather*}
    \caption{\FuncLang's semantics.}
    \label{fig:semantics}
    \Description[Infernce rules for He-Lambda-Small.]
                {Thirteen inference rules defining the semantics of
                He-Lambda-Small choreographies.
                Most of them are entirely normal lambda-calculus rules
                except that they use a mask operator and a specialized
                notion of substitution.
                The exceptions are the COM rules, which use each other
                as recursive preconditions to replace the location annotations
                on unit values.}
    \end{mdframed}
\end{figure}

The \textsc{Com1} rule simply replaces one location-annotation with another.
\textsc{ComPair}, \textsc{ComInl}, and \textsc{ComInr} are defined recursively
amongst each other and \textsc{Com1};
the effect of this is that "data" values can be sent but other values
(functions and variables) cannot.

As is typical for a typed lambda calculus, \FuncLang enjoys preservation and progress.

\begin{theorem}[Preservation]\label{theorem:preservation}
  If $Θ;∅ ⊢ M : T$ and $M \step M'$, then $Θ;∅ ⊢ M' : T$.

  See Appendix~\ref{sec:preservation-proof} for the proof.
\end{theorem}

\begin{theorem}[Progress]\label{theorem:progress}
  If $Θ;∅ ⊢ M : T$, then either M is of form $V$ (which cannot step)
  or their exists $M'$ s.t. $M \step M'$.

  See Appendix~\ref{sec:progress-proof} for the proof.
\end{theorem}

\subsection{The Local Process Language}\label{sec:local-lang}

In order to define EPP and a "ground truth" for \FuncLang computation,
we need a locally-computable language into which it can project.
This local language is very similar to \FuncLang;
to avoid ambiguity we denote local-language expressions $B$ (for "behavior")
instead of $M$ (which denotes a choreographic expression)
and local-language values $L$ instead of $V$.
The syntax is presented in Figure~\ref{fig:local-syntax}.

\begin{figure}[tbhp]
    \begin{mdframed}
    \begin{align*}
        B \BNF   & L \BNFOR B B \BNFOR \CASE{}{B}{x}{B}{x}{B} && \text{Process expressions.} \\
        L \BNF   & x \BNFOR () \BNFOR   λ x \DOT B
                     && \text{Process values.} \\
          \BNFOR & \INL L \BNFOR \INR L \BNFOR  \PAIR L L  \BNFOR  \FST{} \BNFOR \SND{} \\
          \BNFOR & (L, \dots, L) \BNFOR \LOOKUP{n}{} && \text{} \\
          \BNFOR & \RECV{p} \BNFOR \SEND{p^{\ast}}
                     && \text{Receive from one party. Send to many.} \\
          \BNFOR & \SEND{p^{\ast}}^{\ast}
                    && \text{Send to many \emph{and} keep for oneself.} \\
          \BNFOR & ⊥                  && \text{"Missing" (located someplace else).}
    \end{align*}
    \caption{Syntax for a local-process language.}
    \label{fig:local-syntax}
    \Description[A BNF for a simple lambda calculus with "send" and "receive" operators.]
                {A BNF for a simple lambda calculus with "send" and "receive" operators.
                By design, it's very similar to He-Lambda-Small, just without party annotations
                and with "com" replaced by "send", "send*", and "recv".}
    \end{mdframed}
\end{figure}

The local language differs from \FuncLang in a few ways.
It's untyped, and the party-set annotations are mostly missing.
\FuncLang's $\COMM{p}{\nonempty{q}}$ operator is replaced by $\SEND{\nonempty{q}}$ and $\RECV{p}$,
as well as a $\SEND{\nonempty{q}}^{\ast}$, which differs from $\SEND{\nonempty{q}}$ only in that
the process which calls it keeps a copy of the sent value for itself.
Syntactically, the recipient lists of $\SEND{}$ and $\SEND{}^{\ast}$ may be empty;
this keeps semantics consistent in the edge case implied by
a \FuncLang expression like $\COMM{s}{\set{s}}$ (which is useless but legal).
Finally, the value-form ⊥ ("bottom") is a stand-in for parts of the choreography that do not involve the target party.
In the context of choreographic languages, ⊥ does not denote an error but should instead be read as "unknown"
or "somebody else's problem".

The behavior of ⊥ during semantic evaluation can be handled a few different ways,
the pros-and-cons of which are not important in this work.
We use a ⊥-normalizing "floor" function, defined in Figure~\ref{fig:floor},
during EPP and semantic stepping to avoid ever handling
⊥-equivalent expressions like $\PAIR ⊥ ⊥$ or $⊥ ()$.

\begin{figure}[tbhp]
    \begin{mdframed}
\begin{align*}
\FLR{B}                        \DEF      \text{by pattern matching on $B$:}
  & \qquad\qquad\qquad\qquad  \text{\small{(Observe that floor is idempotent.)}} \\
B_1 B_2                      \DEFCASE &
  \begin{cases}
    \FLR{B_1} = ⊥, \FLR{B_2} = L \DEFCASE & ⊥  \\
    \text{else}              \DEFCASE & \FLR{B_1} \FLR{B_2}
  \end{cases}  \\
\CASE{}{B_G}{x_l}{B_l}{x_r}{B_r} \DEFCASE &
  \begin{cases}
    \FLR{B_G} = ⊥                \DEFCASE & ⊥ \\
    \text{else}      \DEFCASE & \CASE{}{\FLR{B_G}}{x_l}{\FLR{B_l}}{x_r}{\FLR{B_r}}
  \end{cases}  \\
λ x \DOT B'                  \DEFCASE & λ x \DOT \FLR{B'} \\
\INL L                       \DEFCASE & \begin{cases}
  \FLR{L} = ⊥                \DEFCASE & ⊥ \\
  \text{else}                \DEFCASE & \INL \FLR{L}
  \end{cases} &&      \\
\INR L                       \DEFCASE & \begin{cases}
  \FLR{L} = ⊥                \DEFCASE & ⊥ \\
  \text{else}                \DEFCASE & \INR \FLR{L}
  \end{cases} &&      \\
\PAIR L_1 L_2                \DEFCASE & \begin{cases}
  \FLR{L_1} = ⊥ = \FLR{L_2}  \DEFCASE & ⊥ \\
  \text{else}                \DEFCASE & \PAIR \FLR{L_1} \FLR{L_2}
  \end{cases} &&         \\
(L_1, \dots, L_n)            \DEFCASE & \begin{cases}
  \forall_{i\in[1,n]} \FLR{L_i} = ⊥ \DEFCASE & ⊥ \\
  \text{else}                \DEFCASE & (\FLR{L_1}, \dots, \FLR{L_n})
  \end{cases} && \\
\begin{rcases}
  x \\
  () \\
  \FST{} \\
  \SND{} \\
  \LOOKUP{i}{} \\
  \SEND{p^{\ast}} \\
  \SEND{p^{\ast}}^{\ast} \\
  \RECV{p} \\
  ⊥
\end{rcases}                 \DEFCASE &  B
\end{align*}
    \caption{The "floor" function, which reduces ⊥-based expressions.}
    \label{fig:floor}
    \Description[A casewise definition of a function using the bottom-brackets associated with a real-number "floor" function.]
                {A casewise definition of a function, using the bottom-brackets associated with a real-number "floor" function,
                that takes a local-language expression and returns it with "bottom" values simplified.}
    \end{mdframed}
\end{figure}

The local semantic stepping rules are given in Figure~\ref{fig:local-semantics}.
Local steps are labeled with send ($⊕$) and receive ($⊖$) sets, like so:
$B \prcstep{\set{(p,L_1), (q,L_2)}}{\set{(r, L_3), (s, L_4)}} B'$,
or $B \prcstep{μ}{η} B'$ when we don't need to inspect the contents of the annotations.
The floor function is used to keep expressions normalized during evaluation.
Otherwise, most of the rules are analogous to the corresponding \FuncLang rules from Figure~\ref{fig:semantics}.
The \textsc{LSend-} rules are defined recursively, similar to the \textsc{Com-} rules.
\textsc{LSendSelf} shows that $\SEND{}^{\ast}$ is exactly like $\SEND{}$
except it locally acts like \inlinecode{id} instead of returning ⊥.
\textsc{LRecv} shows that the $\RECV{}$ operator ignores its argument and can return
\emph{anything}, with the only restriction being that the return value must be reflected in the receive-set step-annotation.

\begin{figure}[tbhp]
    \begin{mdframed}
\begin{gather*}
\myference{LAbsApp}
          {}
          {(λ x \DOT B) L \prcstep{∅}{∅} \FLR{B[x:=L]}}
          \quad
\myference{LApp1}
          {B \prcstep{μ}{η} B'}
          {L B \prcstep{μ}{η} \FLR{L B'}}
          \quad
\myference{LApp2}
          {B \prcstep{μ}{η} B'}
          {B B_2 \prcstep{μ}{η} \FLR{B' B_2}}
          \vdbl
\myference{LCase}
          {B \prcstep{μ}{η} B'}
          {\CASE{}{B}{x_l}{B_l}{x_r}{B_r} \prcstep{μ}{η}
           \FLR{\CASE{}{B'}{x_l}{B_l}{x_r}{B_r}}}
          \vdbl
\myference{LCaseL}
          {}
          {\CASE{}{\INL L}{x_l}{B_l}{x_r}{B_r} \prcstep{∅}{∅} \FLR{B_l[x_l := L]}}
          \quad
\myference{LCaseR}
          {\dots}
          {\dots}
          \vdbl
\myference{LProj1}
          {}
          {\FST{} (\PAIR L_1 L_2) \prcstep{∅}{∅} L_1}
          \quad
\myference{LProj2}
          {\dots}
          {\dots}
          \quad
\myference{LProjN}
          {}
          {\LOOKUP{i}{} (L_1, \dots, L_i, \dots, L_n) \prcstep{∅}{∅} L_i}
          \vdbl
\myference{LSend1}
          {}
          {\SEND{p^{\ast}} () \prcstep{\set{(p, ()) \mid p \in p^{\ast}}}{∅} ⊥}
          \quad
\myference{LSendPair}
          {\SEND{p^{\ast}} L_1 \prcstep{μ_1}{∅} ⊥ \quad
           \SEND{p^{\ast}} L_2 \prcstep{μ_2}{∅} ⊥}
          {\SEND{p^{\ast}} (\PAIR L_1 L_2)
           \prcstep{\set{(p, \PAIR L_1 L_2) \mid p \in p^{\ast}}}{∅}
           ⊥}
          \vdbl
\myference{LSendInL}
          {\SEND{p^{\ast}} L \prcstep{μ}{∅} ⊥}
          {\SEND{p^{\ast}} (\INL L)
           \prcstep{\set{(p, \INL L) \mid p \in p^{\ast}}}{∅}
           ⊥}
          \quad
\myference{LSendInR}
          {\dots}
          {\dots}
          \quad
\myference{LSendSelf}
          {\SEND{p^{\ast}} L \prcstep{μ}{∅} ⊥}
          {\SEND{p^{\ast}}^\ast L \prcstep{μ}{∅} L}
          \vdbl
\myference{LRecv}
          {}
          {\RECV{p} L_0 \prcstep{∅}{\set{(p, L)}} L}
          \quad
\end{gather*}
    \caption{The semantics of the local process language.}
    \label{fig:local-semantics}
    \Description[Fifteen inference rules defining the semantics of the local process language.]
                {Inference rules defining the substitution-based semantic stepping of the local process language.
                There are fifteen rules, roughly corresponding to the similar rules from the choreographic semantics.}
    \end{mdframed}
\end{figure}

\subsection{Endpoint Projection}\label{sec:projection}
Endpoint projection (EPP) is the translation between the choreographic language \FuncLang
and the local process language; necessarily it's parameterized by the specific
local process you're projecting to.
$⟦M⟧_p$ is the projection of $M$ to $p$, as defined in Figure~\ref{fig:eep}.
It does a few things:
Most location annotations are removed, some expressions become ⊥,
⊥-based expressions are normalized by the floor function,
and $\COMM{s}{\nonempty{r}}$ becomes $\SEND{\nonempty{r}}$, $\SEND{\nonempty{r}}^{\ast}$, or $\RECV{s}$,
keeping only the identities of the peer parties and not the local party.

\begin{figure}[tbhp]
    \begin{mdframed}
\begin{align*}
⟦M⟧_p                        \DEF      \text{by pattern matching on $M$:}& \\
N_1 N_2                      \DEFCASE & \FLR{⟦N_1⟧_p ⟦N_2⟧_p} \\
\CASEm{\nonempty{p}}{N}{x_l}{M_l}{x_r}{M_r} \DEFCASE &
  \begin{cases}
    p \in \nonempty{p}       \DEFCASE & \FLR{
      \CASE{}{⟦N⟧_p}{x_l}{⟦M_l⟧_p}{x_r}{⟦M_r⟧_p} } \\
    \text{else}              \DEFCASE & \FLR{
      \CASE{}{⟦N⟧_p}{x_l}{⊥}{x_r}{⊥} }
  \end{cases}  \\
x                            \DEFCASE &  x        \\
(λ x:T \DOT N)@\nonempty{p}  \DEFCASE &
  \begin{cases}
    p \in \nonempty{p}       \DEFCASE & λ x \DOT ⟦N⟧_p \\
    \text{else}              \DEFCASE & ⊥
  \end{cases}  \\
()@\nonempty{p}              \DEFCASE &
  \begin{cases}
    p \in \nonempty{p}       \DEFCASE & () \\
    \text{else}              \DEFCASE & ⊥
  \end{cases}  \\
\INL V                       \DEFCASE & \FLR{\INL ⟦V⟧_p}  &&      \\
\INR V                       \DEFCASE & \FLR{\INR ⟦V⟧_p}  &&      \\
\PAIR V_1 V_2                \DEFCASE & \FLR{\PAIR ⟦V_1⟧_p ⟦V_2⟧_p} &&         \\
(V_1, \dots, V_n)            \DEFCASE & \FLR{(⟦V_1⟧_p, \dots, ⟦V_n⟧_p)} &&       \\
\FST{\nonempty{p}}           \DEFCASE &
  \begin{cases}
    p \in \nonempty{p}       \DEFCASE & \FST{} \\
    \text{else}              \DEFCASE & ⊥
  \end{cases}  \\
\SND{\nonempty{p}}           \DEFCASE &
  \begin{cases}
    p \in \nonempty{p}       \DEFCASE & \SND{} \\
    \text{else}              \DEFCASE & ⊥
  \end{cases}  \\
\LOOKUP{i}{\nonempty{p}}     \DEFCASE &
  \begin{cases}
    p \in \nonempty{p}       \DEFCASE & \LOOKUP{i}{} \\
    \text{else}              \DEFCASE & ⊥
  \end{cases}  \\
\COMM{s}{\nonempty{r}}       \DEFCASE &
  \begin{cases}
    p = s, p \in \nonempty{r}      \DEFCASE & \SEND{\nonempty{r} ∖ \set{p}}^\ast \\
    p = s, p \not\in \nonempty{r}  \DEFCASE & \SEND{\nonempty{r}} \\
    p \not = s, p \in \nonempty{r} \DEFCASE & \RECV{s} \\
    \text{else}              \DEFCASE & ⊥
  \end{cases}
\end{align*}
    \caption{EPP from \FuncLang to the local process language.}
    \label{fig:eep}
    \Description[A casewise definition of a function denoted by double-square-brackets and parameterized by a party-name subscript.]
                {A casewise definition of a function, denoted by enclosing the argument in double-square-brackets,
                parameterized by a party name in subscript,
                that takes a He-Lambda-Small expression and returns the party's view of it in the local process language.}
    \end{mdframed}
\end{figure}

\subsection{Process Networks}\label{sec:networks}
A single party evaluating local code can hardly be considered the ground truth of choreographic computation;
for a message to be sent it must be received \emph{by} someone (and \textit{visa-versa}).
A "network" is a dictionary mapping each party in its domain to a local program representing that party's current place in the execution.
We express party-lookup as $\mathcal{N}(p) = B$.
A singleton network, written $\mathcal{N} = p[B]$, has the one party $p$ in its domain and assigns the expression $B$ to it.
Parallel composition of networks is expressed as $\mathcal{N} \mid \mathcal{N}'$
(the order doesn't matter).
Thus, the following are equivalent:
$\mathcal{N}(p) = B \Longleftrightarrow \mathcal{N} = p[B] \mid \mathcal{N}' \Longleftrightarrow p[B] \in \mathcal{N}$.
When many compositions need to be expressed at once, we can write
$\mathcal{N} = Π_{p \in \nonempty{p}} p[B_p]$.
Parallel projection of all participants in $M$ is expressed as
$⟦M⟧ = Π_{p \in \roles{M}} p[⟦M⟧_p]$.
For example, if $p$ and $q$ are the only parties in $M$, then
$⟦M⟧ = p[⟦M⟧_p] \mid q[⟦M⟧_q]$.

The rules for Network semantics are in Figure~\ref{fig:networks}.
Network semantic steps are annotated with \emph{incomplete} send actions;
$\mathcal{N} \netstep{p}{\set{\dots,(q_i, L_i),\dots}} \mathcal{N}'$
indicates a step in which $p$ sent a respective $L_i$ to each of the listed $q_i$
and the $q_i$s have \emph{not} been noted as receiving.
When there are no such incomplete sends and the $p$ doesn't matter,
it may be omitted for convenience
(\eg $\mathcal{N} \netstep{}{∅} \mathcal{N}'$
instead of $\mathcal{N} \netstep{p}{∅} \mathcal{N}'$).
\textbf{In practice only $∅$-annotated steps are "real".}
Process level semantics only really elevate to network level semantics
when the message-annotations cancel out.
Rule \textsc{NCom} allows annotations to cancel out.
For example the network
$⟦\COMM{s}{\set{p,q}} ()@\set{s}⟧$
gets to $⟦()@\set{p,q}⟧$
by a \emph{single} \textsc{NCom} step.
The derivation tree for that step starts at the top with \textsc{NPro}:
$s[\SEND{\set{p,q}} ()] \netstep{s}{\set{(p,()),(q,())}} s[⊥]$;
this justifies two nestings of \textsc{NCom} in which the $p$ step and $q$ step
(in either order)
compose with the $s$ step and remove the respective party from the step-annotation.

\begin{figure}[tbhp]
    \begin{mdframed}
\begin{gather*}
\myference{NPro}
          {B \prcstep{μ}{∅} B'}
          {p[B] \netstep{p}{μ} p[B']}
          \quad
\myference{NCom}
          {\mathcal{N} \netstep{s}{μ∪\set{(r,L)}} \mathcal{N}'
           \quad B \prcstep{∅}{\set{(s, L)}} B'}
          {\mathcal{N} \mid r[B] \netstep{s}{μ} \mathcal{N}' \mid r[B']}
          \quad
\myference{NPar}
          {\mathcal{N} \netstep{}{∅} \mathcal{N}'}
          {\mathcal{N} \mid \mathcal{N}^{+} \netstep{}{∅} \mathcal{N}' \mid \mathcal{N}^{+}}
\end{gather*}
    \caption{Semantic rules for a network of processes.}
    \label{fig:networks}
    \Description[Inference rules for networks of processes, showing when and how the local semantics can actually be applied.]
                {Three inference rules for networks of processes,
                showing when and how the local semantic stepping rules can be applied in the "real world" of communicating processes.}
    \end{mdframed}
\end{figure}

\subsection{Deadlock Freedom}\label{sec:deadlock-freedom}
Having introduced all of the machinery of EPP and evaluation of a network of communicating processes,
we can now show that the central semantics of \FuncLang is a sound and complete model of that ground truth.

\begin{theorem}[Soundness]\label{theorem:soundness}
  If $Θ;∅ ⊢ M : T$ and $⟦M⟧ \netstep{}{∅}^{\ast} \mathcal{N}_n$,
  then there exists $M'$ such that
  $M \step^{\ast} M'$ and $\mathcal{N}_n \netstep{}{∅}^{\ast} ⟦M'⟧$.

  See Appendix~\ref{sec:soundness-proof} for the proof.
\end{theorem}

\begin{theorem}[Completeness]\label{theorem:completeness}
  If $Θ;∅ ⊢ M : T$ and $M \step M'$,
  then $⟦M⟧ \netstep{}{∅}^{\ast} ⟦M'⟧$.

  See Appendix~\ref{sec:completeness-proof} for the proof.
\end{theorem}

The central promise of choreographic programming is that participants in well-formed choreographies
will never get stuck waiting for messages they never receive.
This important property, \textit{"deadlock freedom by design"}, is trivial once our previous theorems are in place.

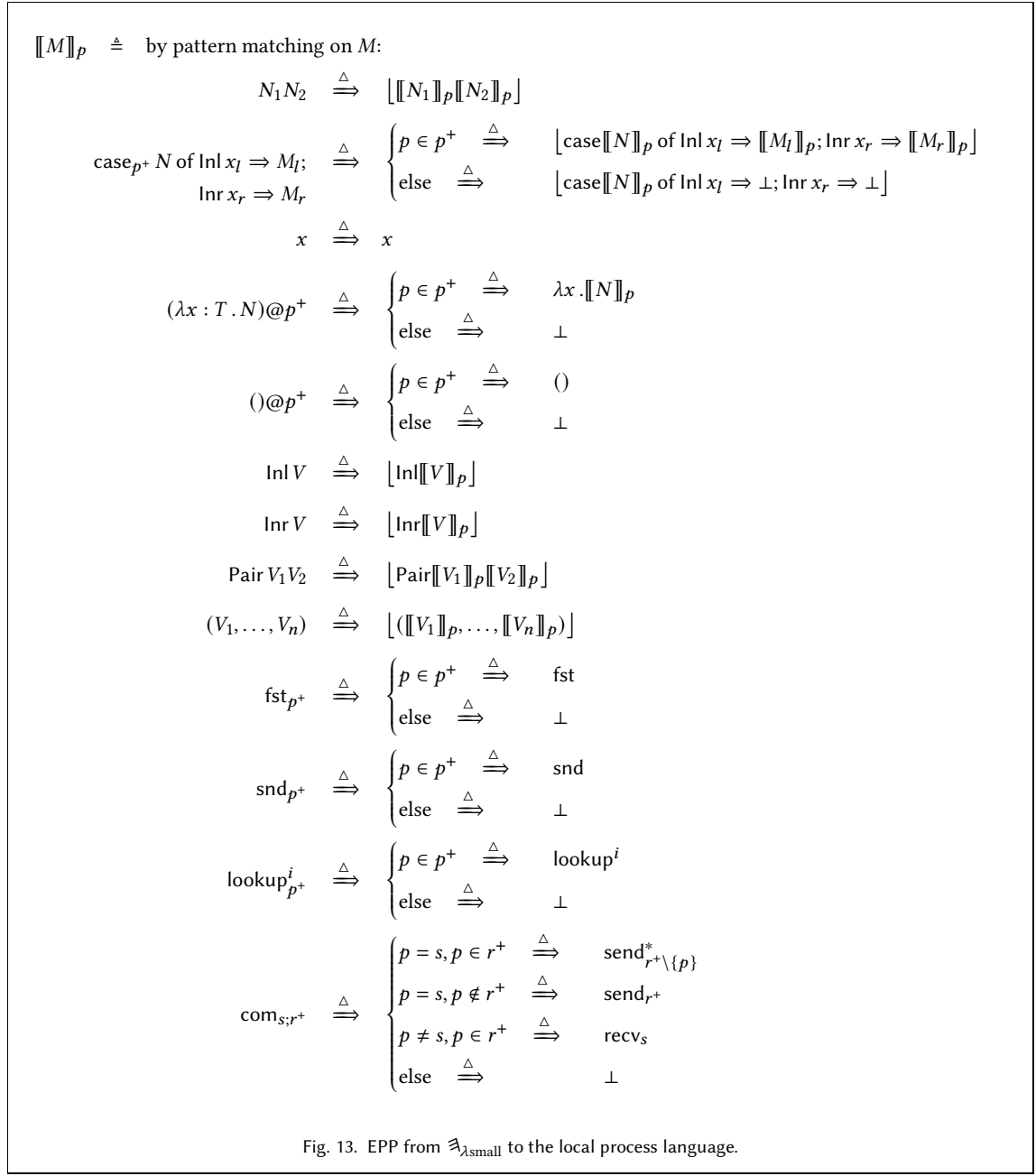
\begin{corollary}[Deadlock Freedom]\label{theorem:deadlock}
  If $Θ;∅ ⊢ M : T$ and $⟦M⟧ \netstep{}{∅}^{\ast} \mathcal{N}$,
    then either $\mathcal{N} \netstep{}{∅}^{\ast} \mathcal{N}'$
    or for every $p\in\roles{M}$, $\mathcal{N}(p)$ is a value.

    This follows from Theorem~\ref{theorem:soundness}, Theorem~\ref{theorem:preservation},
    Theorem~\ref{theorem:progress}, and Theorem~\ref{theorem:completeness}.
\end{corollary}

\section{Case Studies \& Comparisons with Previous Work}\label{sec:comparisons}

A Knowledge of Choice (KoC) strategy is a key component of any safe choreography language.
Any general-purpose KoC strategy will require, at least some of the time,
that parties send messages to each other beyond what would be needed to just to communicate data.
In this section we compare recent choreography languages
to \FuncLang, primarily in terms of how their KoC strategies impact communication efficiency.
By "communication efficiency" we refer to the amount of information sent from each party to each other party
in a choreography that accomplishes some desired global behavior or end state.

For readability, we render \FuncLang examples in this section as plain-text.
To avoid unicode characters, we'll use
\inlinecode{fn} for λ, \inlinecode{=>} for ⇒, \inlinecode{->} for →, and \inlinecode{*} for ×.
The annotations on lambdas, unit, and keyword functions
are given as comma-separated lists in square brackets
(\eg \inlinecode{lookup[2][p_1,p_2,q]} and \inlinecode{com[s][r_1]}).

Furthermore, we sugar our syntax with let-binding,
\eg $(λ var : T \DOT M)@Θ V$ is rendered as \inlinecode{let var : T = V; M},
and often we'll omit the type annotation \inlinecode{T}.
We elide declarations of contextual functions and data types in our examples.
We allow expressions in place of values,
which can be de-sugared to temp variables.
Some of the languages we compare against include polymorphic functions
in their examples;
we annotate such function names in our comparison code,
similar to how our built-ins like \inlinecode{fst} get annotated.

\subsection{HasChor}\label{sec:haschor}
HasChor is a Haskell library for writing choreographies as values
of a monad \inlinecode[haskell]{Choreo}~\cite{haschor}.
Their "just a library" approach, being applied to a mainstream programming language,
limits the safety guarantees they can provide
but is probably necessary for choreographies to see industry use.
The implementation is succinct and easy to use.

HasChor does not have \inlinecode{select} statements;
KoC is handled by broadcasting branch-guards to all participants in the choreography.
This is not efficient.
For example, in Figure~\ref{fig:haschor-kvs} line 6,
it's implicit in the \inlinecode[haskell]{cond} function that
\inlinecode[haskell]{primary} sends the value \inlinecode[haskell]{request'} to everyone
even though \inlinecode[haskell]{client} doesn't need it.
This behavior makes HasChor dangerous to use for any security- or privacy-minded application.
Furthermore, these implicit broadcasts don't bind the data transmitted; it can't be used for anything \emph{besides} KoC.
On line 8 of Figure~\ref{fig:haschor-kvs}, \inlinecode[haskell]{primary} sends \inlinecode[haskell]{backup}
the value \inlinecode[haskell]{request'} \emph{again} so that \inlinecode[haskell]{backup} can actually do work on it.
(In theory it would be possible to recover the bits of information contained in a KoC-only transmission
so that only the one bit of \inlinecode[haskell]{request'} that controls the branching is broadcast
and only the remainder is sent after; doing this in general cases would be substantial work for the user.)
Figure~\ref{fig:our-kvs} shows a more efficient implementation of the same behavior in \FuncLang.

\begin{figure}[tbhp]
    \begin{mdframed}
    \begin{minted}[xleftmargin=10pt,linenos]{haskell}
kvs :: Request @ "client"
       -> (IORef State @ "primary", IORef State @ "backup")
       -> Choreo IO (Response @ "client")
kvs request (primarySt, bkupSt) = do
  request' <- (client, request) ~> primary
  cond (primary, request') \case
    Put _ _ -> do
      req <- (primary, request') ~> backup
      ack <- (backup, \un -> handleRequest (un req) (un bkupSt)) ~~> primary
      return ()
    Get _ -> return ()
  response <- primary `locally` \un -> handleRequest (un request') (un primarySt)
  (primary, response) ~> client
    \end{minted}
    \caption{A HasChor choreography, taken verbatim from \cite{haschor}'s Figure 8.}
    \label{fig:haschor-kvs}
    \Description[A Haskell function called kvs implementing a choreography.]
                {Fourteen lines of Haskell code implementing the interaction between
                a client, a primary server, and a backup server as
                a function from a Request object and the two server-states to a choreography.}
    \end{mdframed}
\end{figure}

\begin{figure}[tbhp]
    \begin{mdframed}
    \begin{minted}[xleftmargin=10pt,linenos]{bash}
(fn request : (PutRequest + GetRequest)@[client] .
  let request_ = com[client][primary] request;
  let req = com[primary][primary, backup] request_;
  let _ = case[primary, backup] req of
    Inl _put => let _ack = com[backup][primary] (handleRequest@[backup] req);
                ()@[primary, backup];
    Inr _get => ()@[primary, backup];
  let response : Response@[primary] = handleRequest@[primary] request_;
  com[primary][client] response
)@[client, primary, backup]
    \end{minted}
    \caption{A \FuncLang choreography implementing the same KVS as in Figure~\ref{fig:haschor-kvs}.}
    \label{fig:our-kvs}
    \Description[An anonymous He-Lambda-Small function.]
                {Ten lines of He-Lambda-Small code implementing the interaction between
                a client, a primary server, and a backup server as
                a function from a Request object to a choreography.}
    \end{mdframed}
\end{figure}

We can also deviate from the structure of the original program to show off
how \FuncLang's multiply-located values enable succinct parallel behavior.
The function in Figure~\ref{fig:our-kvs2} assumes \inlinecode{handleRequest}
relies only on multiply-located state \inlinecode{primary} and \inlinecode{backup} have in common,
and it elides the \inlinecode{_ack} communication.
Whether or not this variation is better would depend on the specific engineering context.

\begin{figure}[tbhp]
    \begin{mdframed}
    \begin{minted}[xleftmargin=10pt,linenos]{bash}
(fn request : (PutRequest + GetRequest)@[client] .
  let req = com[client][primary, backup] request;
  let response : Response@[primary, backup] = handleRequest@[primary, backup] request;
  com[primary][client] response
)@[client, primary, backup]
    \end{minted}
    \caption{A \FuncLang choreography implementing mostly the same behavior as in Figure~\ref{fig:our-kvs}.}
    \label{fig:our-kvs2}
    \Description[An short anonymous He-Lambda-Small function.]
                {Five lines of He-Lambda-Small code implementing the interaction between
                a client, a primary server, and a backup server as
                a function from a Request object to a choreography.}
    \end{mdframed}
\end{figure}

\subsection{ChoRus}\label{sec:chorus}
\cite{chorus} gives a recipe for building a "just a library" choreography system in any modern mainstream language,
and gives an example implementation in Rust: ChoRus.
ChoRus adds two additional operators to the traditional choreography API:
\inlinecode[rust]{enclave} and \inlinecode[rust]{broadcast}.
\inlinecode[rust]{enclave} executes a choreography using a specified sub-universe of parties.
\inlinecode[rust]{broadcast} sends a located value from a specified party to all parties
in the current universe.
In terms of a centralized semantics \inlinecode[rust]{broadcast}'s behavior is to unwrap a located value
into a naked value in the host language;
in Haskell one would express its type as
\inlinecode[haskell]{forall a, (l::Location) . l -> Located l a -> Choreo a}.
This lets ChoRus use the host language's branching operators (\eg \inlinecode{if})
on values generated during choreographic execution.
ChoRus can implement a key-value-store choreography like the ones in Figures~\ref{fig:haschor-kvs} and ~\ref{fig:our-kvs}
with the same communication efficiency as \FuncLang.
The particular pseudo-code example they give is a bookseller protocol shown in Figure~\ref{fig:chorus-2bookseller};
Figure~\ref{fig:our-2bookseller} shows that \FuncLang matches the efficiency of this example too.

\begin{figure}[tbhp]
    \begin{mdframed}
    \begin{minted}[xleftmargin=10pt,linenos]{bash}
two_buyer : Choreo (Option Date @ buyer1)
two_buyer(locally, comm, bcast, enclave) =
  ...
  let decision_buyer1 = locally(buyer1,
                                λ(un) -> un(price_buyer1) ≤ buyer1_budget + contribution)
      in
  let c(locally, comm, bcast, enclave) =
    let decision = bcast(buyer1, decision_buyer1) in
    if decision then
      let delivery_seller = locally(seller,
                                    λ(un) -> catalog.get_delivery(un(title_seller))) in
      let delivery_buyer1 = comm(seller, buyer1, deliver_seller) in
      locally(buyer1, λ(un) -> Some(un(delivery_buyer1)))
    else
      locally(buyer1, λ(un) -> None)
  in enclave([buyer1, seller], c)
    \end{minted}
    \caption{A ChoRus choreography, taken verbatim from \cite{chorus}'s Figure 9.}
    \label{fig:chorus-2bookseller}
    \Description[A ChoRus choreography named "two_buyer"; with the first half elided.]
                {Fifteen lines of ChoRus code, showing the second half of a choreography named "two_buyer".}
    \end{mdframed}
\end{figure}

\begin{figure}[tbhp]
    \begin{mdframed}
    \begin{minted}[xleftmargin=10pt,linenos]{bash}
...
let decision_buyer1 = price_buyer1 ≤ buyer1_budget + contribution;
let decision = com[buyer1][buyer1, seller] decision_buyer;
case[buyer1, seller] decision of
  Inl _ => let delivery_seller = catalog.get_delivery(title);
           let delivery_buyer1 = com[seller][buyer1] deliver_seller;
           Inl delivery_buyer1
  Inr _ => Inr ()@[buyer1]
    \end{minted}
    \caption{A \FuncLang implementation of the choreography in Figure~\ref{fig:chorus-2bookseller}.}
    \label{fig:our-2bookseller}
    \Description[An anonymous He-Lambda-Small choreography implementing the same thing as the previous figure.]
                {Eight lines of He-Lambda-Small code implementing the same "two_buyer" choreography as in the previous figure.}
    \end{mdframed}
\end{figure}

Any \FuncLang lambda induces an enclave, and multicast can be used as broadcast,
so \FuncLang's communication efficiency is at least as good as ChoRus's.
What ChoRus lacks is a way to represent a value that was previously broadcast
to a \emph{sub-universe} of the \emph{current} universe;
in other words, broadcasted-ness is thrown out when exiting an enclave
and all exported values must be singly-located.
Consider the \FuncLang program in Figure~\ref{fig:our-client-server},
in which a server (\inlinecode{carroll}) is ignorant of delegation among two clients.
At \inlinecode{alice}'s direction, she and \inlinecode{bob} agree on either a query of hers or
a query of \inlinecode{bob}'s that she will ask \inlinecode{carroll} to answer.
Note that \inlinecode{bob} only shares his query with \inlinecode{alice} when it's needed,
and \inlinecode{carroll} never knows which query she got.
\inlinecode{carrolls_func} is bound to the variable \inlinecode{answerer} only to give it a type annotation.
\inlinecode{carroll} sends the response to both \inlinecode{alice} and \inlinecode{bob}.
Finally, either \inlinecode{alice} or \inlinecode{bob} run some response-handler function,
depending on the original choice of who's query to use.
ChoRus can represent this choreography approximately,
but introduces extra communication.
In order for \inlinecode{choice} to exist at both \inlinecode{Alice} and \inlinecode{Bob},
it must be \inlinecode{broadcast} inside an enclave.
That means that \inlinecode{choice} is a naked \inlinecode{bool}, and could only leave
the enclave by being wrapped in a (single) location;
in order to have a \inlinecode{choice:bool} variable in scope in \inlinecode{TerminalCho},
a second \inlinecode{broadcast} is needed.
Such an implementation is shown in Figure~\ref{fig:chorus-client-server}, as an excerpt using the ChoRus API.

\begin{figure}[tbhp]
    \begin{mdframed}
    \begin{minted}[xleftmargin=10pt,linenos]{bash}
let choice : ()+()@[alice, bob] = com[alice][alice, bob] alices_choice;
let query : Query@[alice] = case[alice, bob] choice of
  Inl _ => com[bob][alice] bobs_query;
  Inr _ => alices_query;
let answerer : (Query@[carroll] -> Response@[carroll])@[carroll] = carrolls_func;
let response = com[carroll][bob, alice] (answerer (com[alice][carroll] query));
case[alice, bob] choice of
  Inl _ => bobs_terminal response;
  Inr _ => alices_terminal response;
    \end{minted}
    \caption{A \FuncLang implementation of a two-client one-server choreography involving sequential branches.
        Client \inlinecode{bob} may delegate a query against server \inlinecode{carroll},
        or client \inlinecode{alice} may provide the query herself.}
    \label{fig:our-client-server}
    \Description[A He-Lambda-Small choreography with parties "alice", "bob", and "carroll".]
                {A He-Lambda-Small choreography with parties "alice", "bob", and "carroll".
                "carroll" is acting as a server responding to a request chosen amongst "alice" and "bob".}
    \end{mdframed}
\end{figure}

\begin{figure}[tbhp]
    \begin{mdframed}
    \begin{minted}[xleftmargin=10pt,linenos]{rust}
struct MainCho;
impl Choreography for MainCho {
    type L = LocationSet!(Alice, Bob, Carroll);
    fn run(self, op: &impl ChoreoOp<Self::L>) {
        let query = op.enclave(ChooseQueryCho{alices_choice});
        let answerer = op.locally(Carroll, |_| {...});
        let response = op.broadcast(Carroll, op.locally(Carroll, |un| {
            un.unwrap(&answerer)(un.unwrap(&op.comm(Alice, Carroll, &query)))
        }));
        op.enclave(TerminalCho{alices_choice, response});
}}
impl Choreography<Located<String, Alice>> for ChooseQueryCho{
    type L = LocationSet!(Alice, Bob);
    fn run(self, op: &impl ChoreoOp<Self::L>) -> Located<String, Alice> {
        let choice = op.broadcast(Alice, self.alices_choice);
        if choice {
            op.comm(Bob, Alice, &op.locally(Bob, |_|{"Bob?".into()}))
        } else {
            op.locally(Alice, |_|{"Alice?".into()})
        }
}}
impl Choreography for TerminalCho{
    type L = LocationSet!(Alice, Bob);
    fn run(self, op: &impl ChoreoOp<Self::L>) {
        let choice = op.broadcast(Alice, self.alices_choice);
        if choice {
            op.locally(Bob, |un|{un.unwrap(&bobs_terminal)(&self.response)});
        } else {
            op.locally(Alice, |un|{un.unwrap(&alices_terminal)(&self.response)});
        }
}}
    \end{minted}
    \caption{A ChoRus approximation of the client-server-delegation choreography in Figure~\ref{fig:our-client-server}.}
    \label{fig:chorus-client-server}
    \Description[Rust code showing a ChoRus choreography between Alice, Bob, and Carroll.]
                {Rust code showing a ChoRus choreography between Alice, Bob, and Carroll.
                The choreography is almost the same as shown in the previous figure, but there's extra messages sent.}
    \end{mdframed}
\end{figure}

\subsection{Pirouette}\label{pirouette}
Pirouette~\cite{hirsch2021pirouette} is a functional choreographic language.
It uses the \inlinecode{select}-based KoC strategy formalized in \cite{montesi-thesis}:
a branching party sends flag symbols to peers who need to behave differently depending on the branch.
These \inlinecode{select} statements are written explicitly by the user and can be quite parsimonious.
Only if, and not until, the EPPs of the parallel program branches are different for a given user does that user need
to be sent a \inlinecode{select}.
EPP of an \inlinecode{if} statement uses a "merge" operation to combine program branches that are not distinguishable to a given party.
\inlinecode{select} statements project as the \inlinecode{offer} and \inlinecode{choose} operations from multiparty-session-types.

The "merge" function is partial; if needed \inlinecode{select}s are missing from a program
then EPP can fail because the merge of the EPPs of two paths is undefined.
Pirouette's type system doesn't detect this; to check if a Pirouette program is well-formed
one must do all of the relevant endpoint projections.
(All \inlinecode{select}-based systems we've investigated work this way.)
This presents a hurdle against embedding a language like Pirouette as an eDSL in an industrial language like Haskell or Rust:
static analysis of the choreographies cannot be embedded in the host language's type system.
In \cite{hirsch2021pirouette}'s case, they provide a standalone implementation of Pirouette
and Coq proofs of their theorems.

\inlinecode{select} gives good communication efficiency because not every choice needs to be communicated,
but it has some of the limitations of both HasChor and ChoRus.
The \inlinecode{select} flags can't be used as data,
and the Knowledge of Choice they communicate can't be recycled in subsequent conditionals.
To translate our client-server-delegation example from Figure~\ref{fig:our-client-server}
into Pirouette without redundant messages,
the sequential conditionals must be combined and Carroll's part duplicated in each branch.
This is shown in Figure~\ref{fig:pirouette-client-server};
notice that Carroll is never informed which branch she is in; her actions are the same in each case.
In Section~\ref{sec:chor-lambda} we show that \FuncLang's communication efficiency is at-least-as-good as
that of select-and-merge languages.
We believe Pirouette's communication efficiency is at-least-as-good as \FuncLang's,
but scaling the above strategy for combining sequential conditionals across a large codebase
could be challenging.

\begin{figure}[tbhp]
    \begin{mdframed}
    \begin{minted}[xleftmargin=10pt,linenos]{bash}
if alice.choice
  then alice[L] ~> bob;
       bob.bobs_query ~> alice.query;
       alice.query ~> carroll.query;
       carroll.(answerer(query)) ~> bob.response;
       carroll.(answerer(query)) ~> alice.response;
       bob.(terminal response)
  else alice[R] ~> bob;
       alice.alices_query ~> carroll.query;
       carroll.(answerer(query)) ~> bob.response;
       carroll.(answerer(query)) ~> alice.response;
       alice.(terminal response)
    \end{minted}
    \caption{A Pirouette implementation of the client-server-delegation choreography in Figure~\ref{fig:our-client-server}}
    \label{fig:pirouette-client-server}
    \Description[Pirouette code showing a choreography between Alice, Bob, and Carroll.]
                {Pirouette code showing a choreography between Alice, Bob, and Carroll.
                The choreography is almost the same as shown in the previous figure, but re-organized with some repetition.}
    \end{mdframed}
\end{figure}

\subsection{Chorλ}\label{sec:chor-lambda}
Chorλ~\cite{chor-lambda} is a functional choreographic language.
The API and communication efficiency are similar to \cite{hirsch2021pirouette} and \cite{choral},
but \cite{chor-lambda-2} shows that Chorλ's semantics and typing can additionally support structures called \emph{Distributed Choice Types}.
A multiply-located \inlinecode{()@[p,q]} is isomorphic to a tuple of singly-located values \inlinecode{(()@p, ()@q)}.
Distributed Choice Types extend this isomorphism to cover the entire algebra of Unit, Sum, and Product types
in such a way that \inlinecode{p} and \inlinecode{q} never disagree about the value they each have.
Specifically a multiply-located \inlinecode{(A + B)@[p,q]} becomes a singly-located \inlinecode{((A@p, A@q)+(B@p, B@q))},
a type which earlier systems do not support.

Chorλ's "merge" operator supports branching on distributed choice types,
so Chorλ can always match \FuncLang's communication efficiency with a similar program structure
by declaring the needed \inlinecode{multicast[...]} functions.
There are a few disadvantages to writing programs this way:
\begin{itemize}
    \item A distinct \inlinecode{multicast} function needs to be written for every argument-type and every number of recipients.
    \item Functions that compute on singly-located data need to be refactored to unpack data encoded in a distributed-choice-type value.
        Similarly, these new functions would not be generic with respect to the number of parties their arguments were distributed across.
    \item The language still needs to support \inlinecode{select}, so well-formed-ness checking still depends on the partial function "merge"
        (because Chorλ has no other way of implementing the \inlinecode{multicast} functions).
\end{itemize}

Considering the other direction, \FuncLang can likewise match the communication efficiency of Chorλ
and other \inlinecode{select}-based languages.
Typically, this is as simple as multicasting the branch guard to all parties that would have received a \inlinecode{select}
(and to oneself, the original branching party).
Figures~\ref{fig:epp2} and~\ref{fig:epp3} show a simple translation;
in the \FuncLang version the guard-boolean is sent to everyone who was (in the Chorλ version) informed of the choice by \inlinecode{select},
and everyone branches together.
In other situations a party might participate in branches without receiving a \inlinecode{select}
because they don't need to know which one they are in;
this is handled with the reverse of the transformation we showed between Figures~\ref{fig:our-client-server} and~\ref{fig:pirouette-client-server}.

A fully-general algorithmic translation that never compromises on communication efficiency won't maintain the program's structure.
The strategy is as follows:
\begin{itemize}
    \item An expression $M$ involving a party $p$ who doesn't have KoC gets broken into three parts:
        \begin{itemize}
            \item A computation $N_1$ of a cache data structure containing all variables bound up until the first part of $M$ at which $p$ actually does something.
            \item A sub-expression $N_2$ involving $p$. $p$ might be sending a message, receiving a message, receiving a \inlinecode{select}, or doing local computation.
            \item A computation $N_3$ that unpacks the cache from $N_1$ and (possibly) the results from $N_2$ and proceeds with the \emph{continuation}, the remainder of $M$.
                Note that $N_3$ will still need to undergo similar translation.
        \end{itemize}
    \item Since there's KoC that $p$ doesn't have, $M$ must be a branch of a \inlinecode{case}.
        Since the original program was projectable, the other branch must have a similar breakdown
        \emph{with the same $N_2$ middle part}.
        $N_1$, wrapped in a respective $\INL$ or $\INR$, replaces $M$ in the case statement.
        Depending if $N_2$ is to or from $p$, the branches of the new \inlinecode{case} may also have to provide the argument to $N_2$,
        but this should \emph{not} be wrapped in a Sum Type.
    \item If $N_2$ is a \inlinecode{select} operation, then it gets translated into a multicast.
        Its argument, provided by the preceding \inlinecode{case}, will be $\INL ()@\nonempty{q}$ or $\INR ()@\nonempty{q}$ depending on the symbol
                \inlinecode{select}ed\footnote{Chorλ supports arbitrary symbols for \inlinecode{select},
                but since we're concerned with bit-level efficiency we assume the only symbols are \inlinecode{L} and \inlinecode{R}.},
                where $\nonempty{q}$ are the parties who already have KoC.
        Then $\set{p}∪\nonempty{q}$ branch together on the multicast flag.
        The $N_3$ continuations will be handled in duplicate in both of the flag-branches;
        this will often involve dead branches for which applicable caches or behavior do not exist.
        Since these branches will never be hit, it's safe to populate them with default values of the appropriate type.
    \item Otherwise, sequencing of $N_2$ after the $N_1$-generating \inlinecode{case} is straightforward.
    \item To handle the $N_3$ continuations, branch on the cache value (which was wrapped in a Sum Type).
        In each branch, unpack the cached variables (and bind the results of $N_2$ if needed) and proceed with recursive translation of the continuation.
\end{itemize}
Neither \cite{chor-lambda} nor \cite{chor-lambda-2} contain examples requiring such a complicated translation.
Figure~\ref{fig:chor-lambda-complex} shows a made-up Chorλ choreography;
translating it into \FuncLang without compromising communication efficiency is more involved than earlier examples were.
Figure~\ref{fig:our-complex-human} shows how a human might re-implement that choreography in \FuncLang.
Appendix~\ref{sec:translation} contains a more algorithmic translation.

We believe that, while select-\&-merge languages like Chorλ are equivalent
in expressivity and communication efficiency to multi-local-\&-multicast languages like \FuncLang,
\FuncLang's syntax and semantics are more user-friendly for most software engineering purposes.

\begin{figure}[tbhp]
    \begin{mdframed}
    \begin{minted}[xleftmargin=10pt,linenos]{bash}
case ( first_secret[p] ()@p ) of Inl _ => case ( second_secret[p] ()@p ) of
                                            Inl _ => let w = com[q][p] n_q1;
                                                     select[p][q] L;
                                                     let _ = com[p][q] (w + 1@p);
                                                     w + 1@p;
                                            Inr _ => let w = com[q][p] n_q1;
                                                     let y = 2@p;
                                                     select[p][q] L;
                                                     let _ = com[p][q] (w + y);
                                                     w;
                                 Inr _ => let w = com[q][p] n_q1;
                                          case (second_secret[p] ()@p ) of
                                            Inl s => select[p][q] L;
                                                     let _ = com[p][q] 5@p;
                                                     s;
                                            Inr _ => select[p][q] R;
                                                     let z = com[q][p] n_q2;
                                                     w + z;
    \end{minted}
    \caption{A contrived Chorλ choreography that is complicated to efficiently translate into \FuncLang.}
    \label{fig:chor-lambda-complex}
    \Description[An 18-line Chor-Lambda choreography in which there are four parallel branches, only one of which is has special behavior for the second party.]
                {An 18-line Chor-Lambda choreography in which there are four parallel branches
                (a case expression containing a case in each of its two branches).
                Branching is controlled by party "p", and "p" behaves differently in every branch.
                Party "q" has identical behavior in three of the branches, but not the forth.
                In those three branches, "q" is sent the flag "L" via "select", in the forth they are sent "R".
                Thus, although "p" branches on two bits, only one bit needs to be sent to "q" for KoC.}
    \end{mdframed}
\end{figure}

\begin{figure}[tbhp]
    \begin{mdframed}
    \begin{minted}[xleftmargin=10pt,linenos]{bash}
let w = com[q][p] n_q1;
let (cache, flag) = case ( first_secret[p] ()@[p] ) of
  Inl _ => (Inl (second_secret[p] ()@[p]), Inl ()@[p]);
  Inr _ => case (second_secret[p] ()@[p]) of
             Inl s  => (Inr s , Inl ()@[p]);
             Inr s_ => (Inr s_, Inr ()@[p]);  # s_ doesn't get used
let flag_ = com[p][p,q] flag;
case flag_ of Inl _ => let (message, result) = case cache of
                         Inl cl => case cl of
                                     Inl _ => (w + 1@[p], w + 1@[p]);
                                     Inr _ => let y = 2@[p];
                                              (w + y    , w);
                         Inr s => (5@[p], s);
                       let _ = com[p][q] message;
                       result;
              Inr _ => let z = com[q][p] n_q2;
                       w + z
    \end{minted}
        \caption{A \FuncLang re-implementation of the choreography from Figure~\ref{fig:chor-lambda-complex}.}
    \label{fig:our-complex-human}
    \Description[17 lines of He-Lambda-Small code. The program flow from the earlier program largely refactored.]
                {17 lines of He-Lambda-Small code.
                The program flow is greatly different because this is a human translation and
                He-Lambda-Small represents different patterns concisely than Chor-Lambda does.}
    \end{mdframed}
\end{figure}

\section{Related Work}\label{sec:related-work}
Since~\cite{montesi-thesis} formalized the paradigm of choreographic programming,
subsequent work has refined the safety guarantees and relationships with other computational models.
\cite{computationally2015montesi} showed that a small choreography language can
be Turing complete and can be correctly projected to a Turing complete process calculus
while maintaining deadlock freedom.
The same authors followed up more recently with \cite{CRUZFILIPE202038},
where they propose that same language as a canonical model for all choreographic programming.
\cite{giallorenzo_et_al:LIPIcs.ECOOP.2021.22} provide algorithmic translation between
choreographies and multi-tier programs.
\cite{barbanera2022formalisms} shows that some properties of choreographic languages can be
abstracted away from the specifics of any one language's syntax or semantics.
\cite{cruzfilipe2023reasoning} shows that Hoare-style logics can be used to prove functional
correctness properties about choreographies in a \inlinecode{select} based language similar to
\cite{CRUZFILIPE202038}.
\cite{cruzfilipe2023certified} provide a certified compiler to do EPP
on \inlinecode{select}-based choreographies.
\cite{LPAR2023:Keep_me_out_of} explores recursive choreographies using a select-\&-merge
language, but their KoC strategy differs from the languages we examined in
Section~\ref{sec:comparisons} in how it accounts for non-termination.

\paragraph{Diversity of choreographic languages.}
Other work has focused on adding new or alternative language features for choreographies.
\cite{cruzfilipe2017communications} showcased a novel choreographic operation
"multicom", in which an unordered set of communications are represented as simultaneous;
this is more general than "multicast", but would not synergize with multiply-located-values
and doesn't affect KoC.
\cite{graversen2023alice} amends the Chorλ language to make PolyChorλ,
which enjoys polymorphism over both locations and data-types.
\cite{cruzfilipe2023compiles} explore an alternative approach to KoC;
starting with the Core Choreographies language from \cite{CRUZFILIPE202038},
they give a process by which a \emph{non}-well-formed (un-projectable)
choreographic program can be systematically amended into a well-formed one
by adding communication.
\cite{zakhour2023dynamic} augment the notion of a located value with references to values
owned by other parties, and even references to values that are guaranteed to exist but who's
exact location is unknown until runtime.
\cite{plyukhin2024ozone} explores a strategy for out-of-order execution of choreographies;
although their choreographies are written procedurally,
individual parties may evaluate their projections in any order they like (up to data dependencies).

Choral is a JVM-based standalone choreographic language that can interoperate with local Java code~\cite{choral}.
Its communication API is more fine-grained than Pirouette's, but the KoC strategy is the same.
More specifically, directed typed communication channels between parties are objects in Choral,
and parties cannot communicate without access to an appropriate channel.
While this doesn't affect communication efficiency, it does mean that Choral
can be used in contexts where robust communication channels between all parties aren't provided automatically.

Research on choreographies is only beginning to translate into practice.
\cite{eu-business-study} uses implementation of E.U. business regulations
as a case study into the usability of choreographic programming for real-world applications.
\cite{Lugovi__2023} uses the Choral language to implement the IRC online chat system;
notably, their implementation is interoperable with pre-existing clients and servers.

\paragraph{Choreographies in cryptography.}
Meanwhile, as modern cryptographic tools become more complicated and more focused on interacting participants,
researchers in that area have been exploring choreography languages for cryptography.
The only prior instances of choreographic languages with multiply-located values come from applied cryptography.
The .CHO language described in~\cite{bates2024dtsim} is a probabilistic choreographic language with multiply-located values \textit{per se},
but differs from \FuncLang in important ways:
\begin{itemize}
    \item .CHO does not have any branching constructs, so it cannot be described as having any KoC strategy at all.
        There are no choices for the parties to have knowledge \emph{of}.
    \item .CHO is not a higher-order language; it has limited subroutines, but not proper functions.
    \item .CHO is imperative, and builds multiply-located values by transitive \inlinecode{share}ing instead of by multicast.
        \ie instead of a $\COMM{p}{\set{p,q}} V$ (which evaluates to a new value \emph{like} $V$ but with updated location),
        .CHO would say \\ \inlinecode{SEND x TO q}, which makes the pre-existing variable \inlinecode{x}
        available at \inlinecode{q} in addition to wherever it was already located.
\end{itemize}
Although .CHO is an interesting antecedent for multiply-located values, it is not a general-purpose choreography language.

\cite{Sweet_2023}, and previously \cite{wysteria}, construct systems that are remarkably similar to choreographies in their syntax and semantics.
In particular, \cite{Sweet_2023}'s language λ-Symphony has multiply-located values,
\inlinecode{share} and \inlinecode{reveal} functions somewhat similar to multicast,
and their \inlinecode{case} expressions automatically create enclaves.
That said, λ-Symphony is special-purpose for the expression of secure-multiparty computation protocols;
it's dubious if it could be use for other purposes.
\inlinecode{share} encrypts its argument in a special way; the actual data sent to the various recipients is not identical,
and \inlinecode{reveal} requires a similarly encrypted argument which it can decrypt.
The computational model is similar to choreographies, but requires explicit context switching like multi-tier programming.
λ-Symphony is untyped and gives no guarantees that programs won't go wrong in various ways.
Finally, \cite{acay2024secure} use a custom \inlinecode{select}-based choreography language as an intermediate representation for protocol-compilation that ensures cryptographic properties.

\section{Conclusions}\label{sec:conclusion}
We have demonstrated the theoretical soundness and practical ease-of-use of
an alternative core API and accompanying type system and semantics for choreographies.
The \FuncLang language expresses complicated choreographies with efficient communication
and without a specialized operator just for managing Knowledge of Choice.
We have proved that well-typed \FuncLang choreographies never get stuck (in a deadlock or otherwise),
and we have shown by example that \FuncLang choreographies are succinct and easy to reason about.

As part of defining \FuncLang, we formalized the novel choreographic language feature \emph{multiply located values}
(Section~\ref{sec:multiply-located}),
data structures that project (via EPP, Section~\ref{sec:background-epp}) to their own single value at a non-empty set of locations
instead of just one location.
This allows \FuncLang to have an easy-to-use multicast operator instead of only one-to-one communication.
It allows computation that's replicated across a set of locations to be expressed as a single choreographic computation
that doesn't need to be refactored when the number of parties changes.
Finally and most importantly, it reduces the Knowledge of Choice problem into knowledge of data;
participants in a branching expression (\eg \inlinecode{case}) branch together on a guard value they all already possess.
This means that \FuncLang's API doesn't need a \inlinecode{select} operation,
well-formed-ness of choreographies is entirely type-directed,
and EPP doesn't require a partial function for merging branch processes.

We believe that this "multi-local-\&-multicast" style of choreography is more intuitive for new users
than "select-\&-merge" choreographies,
and can implement many real-world protocols more cleanly.
We have shown multiple implementations of several protocols (taken from recent literature and new to demonstrate \FuncLang)
to compare the expressiveness and communication efficiency
of \FuncLang against other recent choreographic languages.
We find that \FuncLang has the same communication efficiency as the best pre-existing languages.
Expressiveness is subjective; we invite the reader to judge that for themselves.
We hope to see multi-local-\&-multicast become a common pattern in choreographic language design and implementation.

\begin{acks}
This material is based upon work supported by the National Science Foundation under Grant No. 2238442 and by the Cold Regions Research and Engineering Laboratory (ERDC-CRREL)
under Contract No. W913E521C0003. Any opinions, findings and conclusions or recommendations expressed in this material are those of the author(s) and do not necessarily reflect the views of the National Science Foundation or the Cold Regions Research and Engineering Laboratory.
\end{acks}

\bibliographystyle{ACM-Reference-Format}
\bibliography{refs}

\appendix

\section{About the language name}\label{sec:name}
We use the Phoenician letter He, written 𐤄, unicode U+10904,
to denote a choreographic language, similar to the way λ denotes a functional language.
The motivation for this choice is that it looks nice;
the justification is that the three lines meeting one connotes collaboration and communication.
𐤄's name, "He", may be pronounced with a "hard e" (to rhyme with "tea") or a "long a" (to rhyme with "bay").
Any non-phonetic connotations it may have had in the Phoenician language
are not a settled matter in archaeology\cite{phoenician1965};
the letter seems to have evolved from an earlier glyph meaning "jubilation",
or joyous celebration\cite{phoenician2010}.
We typeset 𐤄 using the code in Figure~\ref{fig:name}.
The $15^\circ$ tilt is aesthetic; many fonts render 𐤄 that way without such adjustment.

\FuncLang (\inlinecode{He-Lambda-small} where unicode is not available)
is "small" in the sense that it is a parsimonious lambda calculus (and \FuncLang[] doesn't read nicely).
While there's no obvious list of features that would be needed for a "\FuncLang[large\!\!]",
recursion, location-polymorphism, and location-subtyping would certainly be included.

\begin{figure}[tbhp]
    \begin{mdframed}
\begin{minted}[xleftmargin=10pt,linenos]{LaTeX}
\usepackage{newunicodechar}
\usepackage{phoenician}
\newunicodechar{𐤄}{
    \ifmmode{
        \rotatebox[origin=c]{15}{\textphnc{e}}\hspace{-1pt}
    }\else{
        \textphnc{e}
    }\fi}
\end{minted}
    \caption{\LaTeX code for typesetting 𐤄.}
    \label{fig:name}
    \Description[A few lines of LaTeX code.]
                {A few lines of LaTeX code.
                 The phoenician and newunicodechar packages are imported, and newunicodechar "he" is defined
                 with adjustments to its orientation and kerning.}
    \end{mdframed}
\end{figure}

\section{Proof of Theorem Substitution}\label{sec:substitution-proof}

Theorem~\ref{theorem:substitution} says that
if $Θ;Γ,(x:T_x) ⊢ M : T$ and $Θ;Γ ⊢ V : T_x$,
then $Θ;Γ ⊢ M[x := V] : T$.
We first prove a few lemmas.

\begin{lemma}[Enclave]\label{theorem:enclave}
    If $Θ;Γ ⊢ V : T$ and $Θ' \subseteq Θ$
    and $T' = T \mask Θ'$ is defined
    then $V' = V \mask Θ'$ is defined,
    and $Θ';Γ ⊢ V' : T'$.
\end{lemma}

\subsection{Proof of Lemma~\ref{theorem:enclave}}
This is vacuous if $T'$ doesn't exist, so assume it does.
Do induction on the definition of masking for $T$:

\begin{itemize}
\item \textsc{MTData}: $Θ;Γ ⊢ V : d@\nonempty{p}$ and $\nonempty{p} ∩ Θ' ≠ ∅$
  so $T' = d@(\nonempty{p} ∩ Θ')$.
  Consider cases for typing of $V$:
  \begin{itemize}
    \item \textsc{TVar}: $V' = V$ by \textsc{MVVar} and it types by \textsc{TVar} b.c. $T'$ exists.
    \item \textsc{TUnit}: We've already assumed the preconditions for \textsc{MVUnit}, and it types.
    \item \textsc{TPair}: $V = \PAIR V_1 V_2$,
      and $Θ;Γ ⊢ V_1 : d_1@(\nonempty{p_1} \supseteq \nonempty{p})$
      and $Θ;Γ ⊢ V_2 : d_2@(\nonempty{p_2} \supseteq \nonempty{p})$.
      By \textsc{MTData}, these larger-owernership types will still mask with $Θ'$,
      so this case come by induction.
    \item \textsc{TInL}, \textsc{TInR}: Follows by simple induction.
  \end{itemize}
\item \textsc{MTFunction}: $T' = T$ and $\nonempty{p} \subseteq Θ'$,
  so lambdas and function-keywords all project unchanged, and the respective typings hold.
\item \textsc{MTVector}: Simple induction.
\end{itemize}

\begin{lemma}[Quorum]\label{theorem:quorum}
    \textbf{A)} If $Θ;Γ,(x:T_x) ⊢ M : T$ and $T_x' = T_x \mask Θ$, then $Θ;Γ,(x:T_x') ⊢ M : T$.

    \textbf{B)} If $Θ;Γ,(x:T_x) ⊢ M : T$ and $T_x \mask Θ$ is not defined, then $Θ;Γ ⊢ M : T$.
\end{lemma}

\subsection{Proof of Lemma~\ref{theorem:quorum}}
By induction on the typing of M.
The only case that's not recursive or trivial is \textsc{TVar},
for which we just need to observe that masking on a given party-set is idempotent.

\begin{lemma}[Unused]\label{theorem:unused}
  If $Θ;Γ ⊢ M : T$ and $x \not \in Γ$, then $M[x := V] = M$.
\end{lemma}
\subsection{Proof of Lemma~\ref{theorem:unused}}
By induction on the typing of $M$.
There are no non-trivial cases.

\subsection{Proof of Theorem~\ref{theorem:substitution}}
There are 13 cases.
\textsc{TProjN}, \textsc{TProj1}, \textsc{TProj2}, \textsc{TCom}, and \textsc{TUnit}
are trivial base cases.
\textsc{TInL}, \textsc{TInR}, \textsc{TVec}, and \textsc{TPair}
are trivial recursive cases.

\begin{itemize}
  \item \textsc{TLambda} where $T_x' = T_x \mask \nonempty{p}$:
  $M = (λ y : T_y \DOT N)@\nonempty{p}$ and $T = (T_y → T')@\nonempty{p}$.
  \begin{enumerate}
      \item $Θ;Γ,(x:T_x) ⊢ (λ y : T_y \DOT N)@\nonempty{p} : (T_y → T')@\nonempty{p}$ by assumption.
      \item $Θ;Γ ⊢ V : T_x$ by assumption.
      \item $\nonempty{p};Γ,(x:T_x),(y:T_y) ⊢ N : T'$ per preconditions of \textsc{TLambda}.
      \item $Θ;Γ,(y:T_y) ⊢ V : T_x$ by weakening (or strengthening?) \#2.
      \item $V' = V \mask \nonempty{p}$ and $\nonempty{p}; Γ,(y:T_y) ⊢ V' : T_x'$ by Lemma~\ref{theorem:enclave}.
      \item $\nonempty{p};Γ,(x:T_x'),(y:T_y) ⊢ N : T'$ by applying Lemma~\ref{theorem:quorum} to \#3.
      \item $\nonempty{p};Γ,(y:T_y) ⊢ N[x:=V'] : T'$ by induction on \#6 and \#5.
      \item $M[x:=V] = (λ y : T_y \DOT N[x:=V'])@\nonempty{p}$ by definition,
     which typechecks by \#7 and \textsc{TLambda}. \textbf{QED.}
  \end{enumerate}
  \item \textsc{TLambda} where $T_x \mask \nonempty{p}$ is undefined:
  $M = (λ y : T_y \DOT N)@\nonempty{p}$.
  \begin{enumerate}
      \item $\nonempty{p};Γ,(x:T_x),(y:T_y) ⊢ N : T'$ per preconditions of \textsc{TLambda}.
      \item $\nonempty{p};Γ,(y:T_y) ⊢ N : T'$ by Lemma~\ref{theorem:quorum} B.
      \item $N[x:=V] = N$ by Lemma~\ref{theorem:unused},
     so regardless of the existence of $V \mask \nonempty{p}$ the substitution is a noop,
     and it typechecks by \#2 and \textsc{TLambda}.
  \end{enumerate}
  \item \textsc{TVar}: Follows from the relevant definitions, whether $x ≡ y$ or not.
  \item \textsc{TApp}: This is also a simple recursive case;
  the masking of $T_a$ doesn't affect anything.
  \item \textsc{TCase}: Follows the same logic as \textsc{TLambda},
  just duplicated for $M_l$ and $M_r$.
\end{itemize}

\section{Proof of Preservation}\label{sec:preservation-proof}
Theorem~\ref{theorem:preservation} says that
if $Θ;∅ ⊢ M : T$ and $M \step M'$,
then $Θ;∅ ⊢ M' : T$.
We'll need a few lemmas first.

\begin{lemma}[Sub-Mask]\label{theorem:sub-mask}
  If $Θ;Γ ⊢ V : d@\nonempty{p}$ and $∅ ≠ \nonempty{q} \subseteq \nonempty{p}$,
    then \textbf{A:} $d@\nonempty{p} \mask \nonempty{q} = d@\nonempty{q}$ is defined
    and \textbf{B:} $V \mask \nonempty{q}$ is also defined and types as $d@\nonempty{q}$.
\end{lemma}
\subsection{Proof of Lemma~\ref{theorem:sub-mask}}
Part A is obvious by \textsc{MTData}.
Part B follows by induction on the definition of masking for values.
\begin{itemize}
\item \textsc{MVLambda}: Base case; can't happen because it wouldn't allow a data type.
\item \textsc{MVUnit}: Base case; passes definition and typing.
\item \textsc{MVInL}, \textsc{MVInR}: Recursive cases.
\item \textsc{MVPair}: Recursive case.
\item \textsc{MVVector}: Can't happen because it wouldn't allow a data type.
\item \textsc{MVProj1}, \textsc{MVProj2}, \textsc{MVProjN}, and \textsc{MVCom}:
  Base cases, can't happen because they wouldn't allow a data type.
\item \textsc{MVVar}: Base case, trivial.
\end{itemize}

\begin{lemma}[Maskable]\label{theorem:maskable}
  If $Θ;Γ ⊢ V : T$ and $T \mask \nonempty{p} = T'$,
  then \textbf{A:} $V \mask \nonempty{p} = V'$ is defined
    and \textbf{B:} $Θ;Γ ⊢ V' : T'$.
\end{lemma}
\subsection{Proof of Lemma~\ref{theorem:maskable}}
By induction on the definition of masking for values.
\begin{itemize}
\item \textsc{MVLambda}: Base case. From the type-masking assumption, \textsc{MTFunction},
  $\nonempty{p}$ is a superset of the owners,
  so $T' = T$, so $V' = V$.
\item \textsc{MVUnit}: Base case; passes definition and typing.
\item \textsc{MVInL}, \textsc{MVInR}: Recursive cases.
\item \textsc{MVPair}: Recursive case.
\item \textsc{MVVector}: Recursive case.
\item \textsc{MVProj1}, \textsc{MVProj2}, \textsc{MVProjN}, and \textsc{MVCom}:
  From the typing assumption, $\nonempty{p}$ is a superset of the owners,
  so $T' = T$ and $V' = V$.
\item \textsc{MVVar}: Base case, trivial.
\end{itemize}

\begin{lemma}[Exclave]\label{theorem:exclave}
  If $Θ;∅ ⊢ M : T$ and $Θ \subseteq Θ'$
  then $Θ';∅ ⊢ M : T$.
\end{lemma}
\subsection{Proof of Lemma~\ref{theorem:exclave}}
By induction on the typing of $M$.
\begin{itemize}
\item \textsc{TLambda}: The recursive typing is unaffected,
  and the other tests are fine with a larger set.
\item \textsc{TVar}: Can't apply with an empty type context.
\item All other cases are unaffected by the larger party-set.
\end{itemize}

\subsection{Proof of Theorem~\ref{theorem:preservation}}
We prove this by induction on typing rules for $M$.
The eleven base cases (values) fail the assumption that $M$ can step,
so we consider the recursive cases:

\begin{itemize}
\item \textsc{TCase}: $M$ is of form $\CASE{\nonempty{p}}{N}{x_l}{M_l}{x_r}{M_r}$.
  There are three ways it might step:
  \begin{itemize}
  \item \textsc{CaseL}: $N$ is of form $\INL V$, $V'$ exists, and $M' = M_l[x_l := V']$.
    \begin{enumerate}
    \item $\nonempty{p};(x_l:d_l@\nonempty{p}) ⊢ M_l : T$ by the preconditions of \textsc{TCase}.
    \item $Θ;∅ ⊢ V : d_l@\nonempty{p}$ because $N$ must type by \textsc{TInL}.
    \item $\nonempty{p};∅ ⊢ V' : d_l@\nonempty{p}$ by Lemma~\ref{theorem:enclave} and \textsc{MTData}.
    \item $\nonempty{p};∅ ⊢ M_l[x_l := V'] : T$ by Lemma~\ref{theorem:substitution}.
    \item $Θ;∅ ⊢ M_l[x_l := V'] : T$ by Lemma~\ref{theorem:exclave}. \textbf{QED.}
    \end{enumerate}
  \item \textsc{CaseR}: Same as \textsc{CaseL}.
  \item \textsc{Case}: $N \step N'$, and by induction and \textsc{TCase},
    $Θ;Γ⊢ N' : T_N$,
    so the original typing judgment will still apply.
  \end{itemize}
\item \textsc{TApp}: $M$ is of form $F A$, and $F$ is of a function type and $A$ also types
  (both in the empty typing context).
  If the step is by \textsc{App2}or \textsc{App1}, then recursion is easy.
  There are eight other ways the step could happen:
  \begin{itemize}
  \item \textsc{AppAbs}: $F$ must type by \textsc{TLambda}.
    $M = ((λ x : T_x \DOT B)@\nonempty{p}) A$.
    We need to show that $A' = A \mask \nonempty{p}$ exists and $Θ;∅ ⊢ B[x := A'] : T$.
    \begin{enumerate}
    \item $\nonempty{p};(x:T_x) ⊢ B : T$ by the preconditions of \textsc{TLambda}.
    \item $Θ;∅ ⊢ A : T_a'$ such that $T_x = T_a' \mask \nonempty{p}$,
       by the preconditions of \textsc{TApp}.
    \item $A'$ exists and $\nonempty{p};∅ ⊢ A' : T_x$ by Lemma~\ref{theorem:enclave} on \#2.
    \item $\nonempty{p};∅ ⊢ B[x := A'] : T$ by Lemma~\ref{theorem:substitution}.
    \item \textbf{QED.} by Lemma~\ref{theorem:exclave}.
    \end{enumerate}
  \item \textsc{Proj1}: $F = \FST{\nonempty{p}}$ and $A = \PAIR V_1 V_2$ and
    $M' = V_1 \mask \nonempty{p}$.
    Necessarily, by \textsc{TPair} $Θ;∅ ⊢ V_1 : d_1@\nonempty{p_1}$
    where $\nonempty{p} \subseteq \nonempty{p_1}$.
    By Lemma~\ref{theorem:sub-mask}, $Θ;∅ ⊢ M' : T$.
  \item \textsc{Proj2}: same as \textsc{Proj1}.
  \item \textsc{ProjN}: $F = \LOOKUP{i}{\nonempty{p}}$ and $A = (\dots, V_i, \dots)$
    and $M' = V_i \mask \nonempty{p}$.
    Necessarily, by \textsc{TVec} $Θ;∅ ⊢ V_i : T_i$ and $Θ;∅ ⊢ A : (\dots, T_i, \dots)$.
    By \textsc{TApp}, $(\dots, T_i, \dots) \mask \nonempty{p} = T_a$,
    so by \textsc{MTVector} $T_i \mask \nonempty{p}$ exists
    and (again by \textsc{TApp} and \textsc{TProjN}) it must equal $T$.
    \textbf{QED.} by Lemma~\ref{theorem:maskable}.
  \item \textsc{Com1}: By \textsc{TCom} and \textsc{TUnit}.
  \item \textsc{ComPair}: Recusion among the \textsc{Com*} cases.
  \item \textsc{ComInl}:  Recusion among the \textsc{Com*} cases.
  \item \textsc{ComInr}:  Recusion among the \textsc{Com*} cases.
  \end{itemize}
\end{itemize}

\section{Proof of Progress}\label{sec:progress-proof}
Theorem~\ref{theorem:progress} says that
if $Θ;∅ ⊢ M : T$, then either M is of form $V$ (which cannot step)
or their exists $M'$ s.t. $M \step M'$.

The proof is by induction of typing rules.
There are eleven base cases and two recursive cases.
Base cases:
\begin{itemize}
\item \textsc{TLambda}
\item \textsc{TVar} (can't happen, by assumption)
\item \textsc{TUnit}
\item \textsc{TCom}
\item \textsc{TPair}
\item \textsc{TVec}
\item \textsc{TProj1}
\item \textsc{TProj2}
\item \textsc{TProjN}
\item \textsc{TInl}
\item \textsc{TInr}
\end{itemize}

Recursive cases:
\begin{itemize}
\item \textsc{TCase}: $M$ is of form $\CASE{\nonempty{p}}{N}{x_l}{M_l}{x_r}{M_r}$
  and ${Θ;∅ ⊢ N : (d_l + d_r)@\nonempty{p}}$.
  By induction, either $N$ can step, in which case M can step by \textsc{Case},
  or $N$ is a value.
  The only typing rules that would give an $N$ of form $V$ the required type are
  \textsc{TVar} (which isn't compatible with the assumed empty $Γ$),
  and \textsc{TInl} and \textsc{TInr}, which respectively force $N$ to have the required forms
  for $M$ to step by \textsc{CaseL} or \textsc{CaseR}.
  From the typing rules, \textsc{MTData}, and the first part of Lemma~\ref{theorem:enclave},
  the masking required by the step rules is possible.
\item \textsc{TApp}: $M$ is of form $F A$, and $F$ is of a function type and $A$ also types
  (both in the same empty $Γ$).
  By induction, either $F$ can step (so $M$ can step by \textsc{App2}),
  or $A$ can step (so $M$ can step by \textsc{App1}),
  or $F$ and $A$ are both values.
  Ignoring the impossible \textsc{TVar} cases,
  there are five ways an $F$ of form $V$ could type as a function;
  in each case we get to make some assumption about the type of $A$.
  Furthermore, by \textsc{TApp} and Lemma~\ref{theorem:enclave},
  we know that $A$ can mask to the owners of $F$.
  \begin{itemize}
  \item \textsc{TProj1}: $A$ must be a value of type $(d_1×d_2)@\nonempty{q}$,
    and must type by \textsc{TPair}, so it must have form $\PAIR V_1 V_2$,
    so $M$ must step by \textsc{Proj1}.
    We know $V_1$ can mask by \textsc{MVPair}.
  \item \textsc{TProj2}: (same as \textsc{TProj1})
  \item \textsc{TProjN}: $A$ must be a value of type $(T_1,\dots,T_n)$ with $i ≤ n$
    and must type by \textsc{TVec}, so it must have from $(V_1,\dots,V_n)$.
    $M$ must step by \textsc{ProjN}.
    We known $V_i$ can step by \textsc{MVVector}.
  \item \textsc{TCom}: $A$ must be a value of type $d@\nonempty{q}$,
      such that $d@\nonempty{q} \mask \nonempty{s} = d@\nonempty{s}$.
          For that to be true, \textsc{MTData} requires that $\nonempty{s} \subseteq \nonempty{q}$.
    $A$ can type that way under \textsc{TUnit}, \textsc{TPair}, \textsc{TInl}, or \textsc{TInr},
    which respectively force forms $()@\nonempty{q}$, $\PAIR V_1 V_2$, $\INL V$, and $\INR V$,
    which respectively require that $M$ reduce by
    \textsc{Com1}, \textsc{ComPair}, \textsc{ComInl}, and \textsc{ComInr}.
          In the case of $()$, this follows from Lemma~\ref{theorem:sub-mask},
          since $\set{s} \subseteq \nonempty{s} \subseteq \nonempty{q}$;
    the other three are recursive among each other.
  \item \textsc{TLambda}: $M$ must reduce by \textsc{AppAbs}.
      By the assumption of \textsc{TApp} and Lemma~\ref{theorem:maskable}, it can.
  \end{itemize}
\end{itemize}

\section{Proof of Theorem Soundness}\label{sec:soundness-proof}
Theorem~\ref{theorem:soundness} says that
if $Θ;∅ ⊢ M : T$ and $⟦M⟧ \netstep{}{∅}^{\ast} \mathcal{N}_n$,
then there exists $M'$ such that
$M \step^{\ast} M'$ and $\mathcal{N}_n \netstep{}{∅}^{\ast} ⟦M'⟧$.
We'll need a few lemmas first.

\begin{lemma}[Values]\label{theorem:values}
  \textbf{A):} $⟦V⟧_p = L$.
  \textbf{B):} If $⟦M⟧_p = L \neq ⊥$ then $M$ is a value $V$.

  Proof is by inspection of the definition of projection.
\end{lemma}
\begin{corollary}\label{theorem:values-cor}
  If $N$ is well-typed and $⟦N⟧$ can step at all,
    then \textbf{(A)} $N$ can step to some $N'$
    and \textbf{(B)} $⟦N⟧$ can multi-step to $⟦N'⟧$ with empty annotation.

    \textbf{A} follows from Lemma~\ref{theorem:values} and Theorem~\ref{theorem:progress}.
    \textbf{B} is just Theorem~\ref{theorem:completeness}.
\end{corollary}

\begin{lemma}[Determinism]\label{theorem:determinism}
  If
  $\mathcal{N}_a \mid \mathcal{N}_0 \netstep{}{∅} \mathcal{N}_a \mid \mathcal{N}_1$
  s.t. for every $p[B_0] \in \mathcal{N}_0$, $\mathcal{N}_1(p) \neq B_0$, \\
    \emph{and}
  $\mathcal{N}_b \mid \mathcal{N}_0 \netstep{}{∅} \mathcal{N}_c \mid \mathcal{N}_2$
  s.t. the domain of $\mathcal{N}_2$ equals the domain of $\mathcal{N}_0$,  
    then \emph{either}
    \begin{itemize}
        \item $\mathcal{N}_2 = \mathcal{N}_0$, \emph{or}
        \item $\mathcal{N}_2 = \mathcal{N}_1$ and $\mathcal{N}_b = \mathcal{N}_c$.
    \end{itemize}
\end{lemma}

\subsection{Proof of Lemma~\ref{theorem:determinism}} First, observe that for every non-value expression in the process language,
there is at most one rule in the process semantics by which it can step.
(For values, there are zero.)
Furthermore, the only way for
the step annotation and resulting expression to \emph{not} be fully determined
by the initial expression
is if the justification is based on a \textsc{LRecv} step,
in which case the send-annotation will be empty
and the resulting expression will match the (single) item in the receive-annotation.

$\mathcal{N}_a \mid \mathcal{N}_0 \netstep{}{∅} \mathcal{N}_a \mid \mathcal{N}_1$
must happen by \textsc{NPar}, so consider the $\mathcal{N}_0$ step that enables it;
call that step \stepname{S}.
\stepname{S} can't be by \textsc{NPar};
that would imply parties in $\mathcal{N}_0$ who don't step.
\begin{itemize}
    \item If \stepname{S} is by \textsc{NPro}, then $\mathcal{N}_0 = p[B_0]$ is a singleton
  and \stepname{S} is justified by a process step with empty annotation.
  As noted above, that process step is the only step $B_0$ can take,
  so the
  $\mathcal{N}_b \mid \mathcal{N}_0 \netstep{}{∅} \mathcal{N}_c \mid \mathcal{N}_2$
  step must either be a \textsc{NPar} composing some other party(ies) step
  with $\mathcal{N}_0$ (satisfying the first choice),
  or a \textsc{NPar} composing \stepname{S} with $\mathcal{N}_b$
  (satisfying the second).
\item If \stepname{S} is by \textsc{NCom}, then there must be both
  a singleton \textsc{NPro} step justified by a process step
  (by some party $s$)
  with nonempty send-annotation
  and a nonempty sequence of other party steps
  (covering the rest of $\mathcal{N}_0$'s domain)
  that it gets matched with
  each with a corresponding receive-annotation.
  The send-annotated \textsc{NPro} step is deterministic in the same way as
  an empty-annotated \textsc{NPro} step.
  In order for the parties to cancel out, it can only compose by \textsc{NCom}
  with (a permutation of) the same sequence of peers.
  Considered in isolation, the peers are non-deterministic,
  but their process-steps can only be used in the network semantics by composing
  with $s$ via \textsc{NCom},
  and their resulting expressions are determined by the matched process annotation,
  which is determined by $s$'s step. \\
  Thus, for any $p[B_2] \in \mathcal{N}_2$,
  $B_2 \neq \mathcal{N}_0(p)$ implies that
  for all $q[B_2'] \in \mathcal{N}_2$, $B_2' = \mathcal{N}_1(p)$.
  In the case where $\mathcal{N}_2 = \mathcal{N}_1$,
  the step from $\mathcal{N}_0$ could only have composed with
  $\mathcal{N}_b$ by \textsc{NPar},
  so $\mathcal{N}_b = \mathcal{N}_c$, Q.E.D.
\end{itemize}

\begin{lemma}[Parallelism]\label{theorem:parallelism}
  \textbf{A):} If $\mathcal{N}_1 \netstep{}{∅}^{\ast} \mathcal{N}_1'$
  and $\mathcal{N}_2 \netstep{}{∅}^{\ast} \mathcal{N}_2'$
  then $\mathcal{N}_1 \mid \mathcal{N}_2 \netstep{}{∅}^{\ast}
  \mathcal{N}_1' \mid \mathcal{N}_2 \netstep{}{∅}^{\ast}
  \mathcal{N}_1' \mid \mathcal{N}_2'$. \\
  \textbf{B):} If $\mathcal{N}_1 \mid \mathcal{N}_2 \netstep{}{∅}^{\ast}
  \mathcal{N}_1' \mid \mathcal{N}_2 \netstep{}{∅}^{\ast}
  \mathcal{N}_1' \mid \mathcal{N}_2'$,
  then $\mathcal{N}_1 \netstep{}{∅}^{\ast} \mathcal{N}_1'$
  and $\mathcal{N}_2 \netstep{}{∅}^{\ast} \mathcal{N}_2'$.
\end{lemma}

\subsection{Proof of Lemma~\ref{theorem:parallelism}}
\textbf{A} is just repeated application of \textsc{NPar}. \\
For \textbf{B}, observer that in the derivation tree of ever step of the sequence, some (possibly different)
minimal sub-network will step by \textsc{NPro} or {NCom} as a precondition
to some number of layers of \textsc{NPar}.
The domains of these minimal sub-networks will be subsets of the domains of $\mathcal{N}_1$
and $\mathcal{N}_2$ respectively,
so they can just combine via \textsc{NPar} to get the needed step in the respective sequences for
$\mathcal{N}_1$ and $\mathcal{N}_2$.

\subsection{Proof of Theorem~\ref{theorem:soundness}}
Declare the predicate $\mathsf{sound}(\mathcal{N})$ to mean that
there exists some $M_{\mathcal{N}}$ such that
$M \step^{\ast} M_{\mathcal{N}}$
and $\mathcal{N} \netstep{}{∅}^{\ast} ⟦M_{\mathcal{N}}⟧$.

Consider the sequence of network steps
$⟦M⟧ = \mathcal{N}_0 \netstep{}{∅} \dots \netstep{}{∅} \mathcal{N}_n$.
By Corollary~\ref{theorem:values-cor}, $\mathsf{sound}(\mathcal{N}_0)$.
Select the largest $i$ s.t. $\mathsf{sound}(\mathcal{N}_i)$.
We will derive a contradiction from an assumption that
$\mathcal{N}_{i+1}$ is part of the sequence;
this will prove that $i=n$, which completes the proof of the Theorem.

Choose a sequence of network steps (of the possibly many such options)
$\mathcal{N}_i = \mathcal{N}^a_i \netstep{}{∅} \dots \netstep{}{∅}
\mathcal{N}^a_m = ⟦M^a⟧$
where $M \step^{\ast} M^a$.

Assume $\mathcal{N}_{i+1}$ is part of the original sequence.
Decompose the step to it as
$\mathcal{N}_i = \mathcal{N}^0_i \mid \mathcal{N}^1_i \netstep{}{∅}
\mathcal{N}^0_i \mid \mathcal{N}^1_{i+1} = \mathcal{N}_{i+1}$
where $\mathcal{N}^1_i$'s domain is as large as possible.
We will examine two cases:
either the parties in $\mathcal{N}^1_i$ make steps in the sequence to
$\mathcal{N}^a_m$, or they do not.
Specifically, consider the largest $j$ s.t.
$\mathcal{N}^a_j = \mathcal{N}^b_j \mid \mathcal{N}^1_i$.

\begin{itemize}
\item Suppose $j < m$. \\
  By Lemma~\ref{theorem:determinism} and our decision that $j$ is as large as possible,
  $\mathcal{N}^a_{j+1} = \mathcal{N}^b_j \mid \mathcal{N}^1_{i+1}$.
  Thus we have
  $\mathcal{N}^0_i \mid \mathcal{N}^1_i \netstep{}{∅}^{\ast}
   \mathcal{N}^b_j \mid \mathcal{N}^1_i \netstep{}{∅}
   \mathcal{N}^b_j \mid \mathcal{N}^1_{i+1}$.
  By Lemma~\ref{theorem:parallelism}, we can reorganize that into an alternative sequence where
  $\mathcal{N}^0_i \mid \mathcal{N}^1_i \netstep{}{∅}
   \mathcal{N}^0_i \mid \mathcal{N}^1_{i+1} \netstep{}{∅}^{\ast}
   \mathcal{N}^b_j \mid \mathcal{N}^1_{i+1}$.
  Since $\mathcal{N}^0_i \mid \mathcal{N}^1_{i+1} = \mathcal{N}_{i+1}$
  and $\mathcal{N}^a_{j+1} \netstep{}{∅}^{\ast} ⟦M^a⟧$,
  this contradicts our choice that $i$ be as large as possible.
\item Suppose $j = m$, so $⟦M^a⟧ = \mathcal{N}^b_m \mid \mathcal{N}^1_i$.\\
  By Lemma~\ref{theorem:parallelism}, $⟦M^a⟧$ can step (because $\mathcal{N}^1_i$ can step)
  so by Corollary~\ref{theorem:values-cor}, $M^a \step M^{a+1}$.
  We can repeat our steps from our choice of
  $\mathcal{N}^a_i \netstep{}{∅}^{\ast} \mathcal{N}^a_m = ⟦M^a⟧$,
  but using $M^{a+1}$ instead of $M^a$.
        Since \FuncLang doesn't have recursion, eventually we'll arrive at a $M^{a++}$
  that can't step, and then-or-sooner we'll be in the first case above.
  Q.E.D.
\end{itemize}

\section{Proof of Theorem Completeness}\label{sec:completeness-proof}
Theorem~\ref{theorem:completeness} says that
if $Θ;∅ ⊢ M : T$ and $M \step M'$,
then $⟦M⟧ \netstep{}{∅}^{\ast} ⟦M'⟧$.
We'll need a few lemmas first.

\begin{lemma}[Cruft]\label{theorem:cruft}
  If $Θ;∅ ⊢ M : T$ and $p \not\in Θ$,
  then $⟦M⟧_p = ⊥$.
\end{lemma}
\subsection{Proof of Lemma~\ref{theorem:cruft}}
By induction on the typing of $M$:
\begin{itemize}
\item \textsc{TLambda}:
  $\nonempty{p} \subseteq Θ$, therefore $p \not\in \nonempty{p}$,
  therefore $⟦M⟧_p = ⊥$.
\item \textsc{TVar}: Can't happen because $M$ types with empty $Γ$.
\item \textsc{TUnit}, \textsc{TCom}, \textsc{TProj1}, \textsc{TProj2},
  and \textsc{TProjN}:
  Same as \textsc{TLambda}.
\item \textsc{TPair}, \textsc{TVec}, \textsc{TInl}, and \textsc{TInr}:
  In each of these cases we have some number of recursive typing judgments
  to which we can apply the inductive hypothesis.
  This enables the respective cases of the definition of floor
  (as used in the respective cases of the definition of projection)
  to map to $⊥$.
\item \textsc{TApp}: $M = N_1 N_2$.
  By induction, $⟦N_1⟧_p = ⊥$ and $⟦N_2⟧_p = ⊥$,
  so $⟦M⟧_p = ⊥$
\item \textsc{TCase}: Similar to \textsc{TLambda},
  by induction the guard projects to $⊥$ and therefore the whole thing does too.
\end{itemize}

\begin{lemma}[Existence]\label{theorem:existence}
  If $Θ;Γ ⊢ V : d@\nonempty{p}$ and $p,q \in \nonempty{p}$,
  then $⟦V⟧_p = ⟦V⟧_q \neq ⊥$.
\end{lemma}
\subsection{Proof of Lemma~\ref{theorem:existence}}
By induction on possible typings of $V$:
\begin{itemize}
\item \textsc{TVar}: Projection is a no-op on variables.
\item \textsc{TUnit}: $⟦V⟧_p = ⟦V⟧_q = ()$.
\item \textsc{TPair}: $p,q \in \nonempty{p_1} ∩ \nonempty{p_2}$,
  so both are in each of them, so we can recurse on $V_1$ and $V_2$.
\item \textsc{TInl} and \textsc{TInr}: simple induction.
\end{itemize}

\begin{lemma}[Bottom]\label{theorem:bottom}
  If $Θ;∅ ⊢ M : T$ and $⟦M⟧_p = ⊥$ and $M \step M'$
  then $⟦M'⟧_p = ⊥$.
\end{lemma}
\subsection{Proof of Lemma~\ref{theorem:bottom}}
By induction on the step $M \step M'$.
\begin{itemize}
\item \textsc{AppAbs}: $M = (λ x:T_x \DOT N)@\nonempty{p} V$,
  and necessarily $⟦(λ x:T_x \DOT N)@\nonempty{p}⟧_p = ⊥$.
  Since the lambda doesn't project to a lambda, $p\not\in\nonempty{p}$.
  $M' = N[x:=V\mask\nonempty{p}]$.
        By \textsc{TLambda}, Lemma~\ref{theorem:substitution}, and Lemma~\ref{theorem:cruft},
  $⟦N[x:=V\mask\nonempty{p}]⟧_p = ⊥$.
\item \textsc{App1}: $M = V N$
  and necessarily $⟦V⟧_p = ⟦N⟧_p = ⊥$.
  By induction on $N \step N'$, $⟦N'⟧_p = ⊥$.
\item \textsc{App2}: Same as \textsc{App1}.
\item \textsc{Case}: The guard must project to $⊥$, so this follows from induction.
\item \textsc{CaseL} (and \textsc{CaseR} by mirror image):
  $M = \CASE{\nonempty{p}}{\INL V}{x_l}{M_l}{x_r}{M_r}$
  and $M' = M_l[x_l := V\mask\nonempty{p}]$.
  Necessarily, $⟦V⟧_p = ⊥$.
  By \textsc{TCase} and \textsc{MTData}, $\INL V$ types as data,
        so by Lemma~\ref{theorem:existence} $p \not\in \nonempty{p}$.
        By \textsc{TCase}, Lemma~\ref{theorem:substitution}, and Lemma~\ref{theorem:cruft},
  $⟦M'⟧_p = ⟦M_l[x_l := V\mask\nonempty{p}]⟧_p = ⊥$.
\item \textsc{Proj1}: $M = \FST{\nonempty{p}}(\PAIR V_1 V_2)$,
  and $p \not \in \nonempty{p}$.
  $M' = V_1 \mask \nonempty{p}$.
  Since $Θ;∅ ⊢ V_1 : T'$ (by \textsc{TPair})
  and $T' \mask \nonempty{p} = T''$ is defined
  (by \textsc{TApp} and the indifference of \textsc{MTData} to the data's structure),
        by Lemma~\ref{theorem:enclave} $\nonempty{p};∅ ⊢ V_1 \mask \nonempty{p} : T''$.
        By Lemma~\ref{theorem:cruft} this projects to $⊥$.
\item \textsc{Proj2}, \textsc{ProjN}, and \textsc{Com1} are each pretty similar to
  \textsc{Proj1}.
\item \textsc{Com1}, \textsc{ComPair}, \textsc{ComInl}, and \textsc{ComInr}:
    For $M$ to project to ⊥, $p$ must be neither a sender nor a recipient.
    By induction among these cases (with \textsc{Com1} as the base case),
        $M'$ will be some structure of $()@\nonempty{r}$;
        since $p\not\in\nonempty{r}$ and projection uses floor,
        this will project to ⊥.
\end{itemize}

\begin{lemma}[Masked]\label{theorem:masked}
  If $p \in \nonempty{p}$ and $V' = V \mask \nonempty{p}$
  then $⟦V⟧_p = ⟦V'⟧_p$.
\end{lemma}
\subsection{Proof of Lemma~\ref{theorem:masked}}
By (inductive) case analysis of endpoint projection:
\begin{itemize}
\item $⟦x⟧_p = x$. By \textsc{MVVar} the mask does nothing.
\item $⟦(λ x:T \DOT M)@\nonempty{q}⟧_p$:
  Since $V \mask \nonempty{p}$ is defined, by \textsc{MVLambda} it does nothing.
\item $⟦()@\nonempty{q}⟧_p$: By \textsc{MVUnit} $V' = ()@(\nonempty{p} ∩ \nonempty{q})$.
  $p$ is in that intersection iff $p \in \nonempty{q}$,
  so the projections will both be $()$ or $⊥$ correctly.
\item $\INL V_l$, $\INR V_r$, $\PAIR V_1 V_2$, $(V_1, \dots, V_n)$: simple recursion.
\item $\FST{\nonempty{q}}$, $\SND{\nonempty{q}}$,
  $\LOOKUP{i}{\nonempty{q}}$, $\COMM{q}{\nonempty{q}}$:
  Since the masking is defined, it does nothing.
\end{itemize}

\begin{lemma}[Floor Zero]\label{theorem:floor-zero}
  $⟦M⟧_p = \FLR{⟦M⟧_p}$
\end{lemma}
\subsection{Proof of Lemma~\ref{theorem:floor-zero}}
There are thirteen forms.
Six of them (application, case, injection-r/l, pair and vector)
apply floor directly in the definition of projection.
Six of them (variable, unit, the three lookups, and $\langword{com}$)
can only project to values such that floor is a no-op.
For a lambda $(λ x:T_x \DOT N)@\nonempty{p}$, the proof is by induction on the body $N$.

\begin{lemma}[Distributive Substitution]\label{theorem:distributive-substitution}
  If $Θ;(x : T_x) ⊢ M : T$ and $p \in Θ$, \\
  then $⟦M[x:=V]⟧_p = \FLR{⟦M⟧_p[x := ⟦V⟧_p]}$.
    (Because $⟦V⟧_p$ may be ⊥, this isn't really distribution; an extra flooring operation is necessary.)
\end{lemma}
\subsection{Proof of Lemma~\ref{theorem:distributive-substitution}}
It'd be more elegant if substitution really did distribute over projection,
but this weaker statement is what we really need anyway.
The proof is by inductive case analysis on the form of $M$:
\begin{itemize}
\item $\PAIR V_1 V_2$: $⟦M[x:=V]⟧_p = ⟦\PAIR V_1[x:=V] V_2[x:=V]⟧_p
  = \FLR{\PAIR ⟦V_1[x:=V⟧_p ⟦V_2[x:=V]⟧_p}$ \\
  and $⟦M⟧_p[x := ⟦V⟧_p] = \FLR{\PAIR ⟦V_1⟧_p ⟦V_2⟧_p}[x := ⟦V⟧_p]$.
  \begin{itemize}
  \item Suppose one of $⟦V_1⟧_p$, $⟦V_2⟧_p$ is not $⊥$.
    Then \\
    $⟦M⟧_p[x := ⟦V⟧_p] = (\PAIR \FLR{⟦V_1⟧_p} \FLR{⟦V_2⟧_p})[x := ⟦V⟧_p]$ \\
          which by Lemma~\ref{theorem:floor-zero}
    $= (\PAIR ⟦V_1⟧_p ⟦V_2⟧_p)[x := ⟦V⟧_p]
     = \PAIR (⟦V_1⟧_p[x := ⟦V⟧_p]) (⟦V_2⟧_p[x := ⟦V⟧_p])$. \\
    Thus $\FLR{⟦M⟧_p[x := ⟦V⟧_p]}
     = \FLR{\PAIR (⟦V_1⟧_p[x := ⟦V⟧_p]) (⟦V_2⟧_p[x := ⟦V⟧_p])}$. \\
    By induction,
    $⟦V_1[x:=V]⟧_p = \FLR{⟦V_1⟧_p[x := ⟦V⟧_p]}$ and
    $⟦V_2[x:=V]⟧_p = \FLR{⟦V_2⟧_p[x := ⟦V⟧_p]}$;
    with that in mind,
    \begin{itemize}
    \item Suppose one of $⟦V_1[x:=V]⟧_p$, $⟦V_1[x:=V]⟧_p$ is not $⊥$. \\
      $\FLR{⟦M⟧_p[x := ⟦V⟧_p]}
       = \PAIR \FLR{⟦V_1⟧_p[x := ⟦V⟧_p]} \FLR{⟦V_2⟧_p[x := ⟦V⟧_p]}$, \\
      and $⟦M[x:=V]⟧_p = \PAIR \FLR{⟦V_1[x:=V⟧_p} \FLR{⟦V_2[x:=V]⟧_p}
       = \PAIR ⟦V_1[x:=V⟧_p ⟦V_2[x:=V]⟧_p$
      Q.E.D.
    \item Otherwise, $\FLR{⟦M⟧_p[x := ⟦V⟧_p]} = ⊥ = ⟦M[x:=V]⟧_p$.
    \end{itemize}
  \item Otherwise, $⟦M⟧_p[x := ⟦V⟧_p] = \FLR{\PAIR ⊥ ⊥}[x := ⟦V⟧_p] = ⊥$. \\
      Note that, by induction \textit{etc},
    $⟦V_1⟧_p = ⊥ = ⟦V_1⟧_p[x := ⟦V⟧_p] = \FLR{⟦V_1⟧_p[x := ⟦V⟧_p]}
     = ⟦V_1[x:=V]⟧_p$,
    and the same for $V_2$, so
    $⟦M[x:=V]⟧_p = ⊥$, Q.E.D.
  \end{itemize}
\item $\INL V_l$, $\INR V_r$, $(V_1, \dots, V_n)$:
  Follow the same inductive pattern as $\PAIR$.
\item $N_1 N_2$:
  $⟦M[x:=V]⟧_p = ⟦N_1[x:=V] N_2[x:=V]⟧_p = \FLR{⟦N_1[x:=V]⟧_p ⟦N_2[x:=V]⟧_p}$ \\
  $= \begin{cases}
    \FLR{⟦N_1[x:=V]⟧_p} = ⊥, \FLR{⟦N_2[x:=V]⟧_p} = L :& ⊥ \\
    \text{else} :& \FLR{⟦N_1[x:=V]⟧_p} \FLR{⟦N_2[x:=V]⟧_p}
  \end{cases}$ \\
  $= \begin{cases}
    ⟦N_1[x:=V]⟧_p = ⊥, ⟦N_2[x:=V]⟧_p = L :& ⊥ \\
    \text{else} :& ⟦N_1[x:=V]⟧_p ⟦N_2[x:=V]⟧_p
  \end{cases}$ \\
  and $\FLR{⟦M⟧_p[x:=⟦V⟧_p]} = \FLR{\FLR{⟦N_1⟧_p ⟦N_2⟧_p}[x:=⟦V⟧_p]}$ \\
  $= \begin{cases}
    \FLR{⟦N_1⟧_p} = ⊥, \FLR{⟦N_2⟧_p} = L :& \FLR{⊥[x:=⟦V⟧_p]} = ⊥ \\
    \text{else} :& \FLR{ (\FLR{⟦N_1⟧_p} \FLR{⟦N_2⟧_p})[x:=⟦V⟧_p] } \\
                 & \quad= \FLR{ (⟦N_1⟧_p[x:=⟦V⟧_p]) (⟦N_2⟧_p[x:=⟦V⟧_p]) }
  \end{cases}$ \\
  $= \begin{cases}
    \FLR{⟦N_1⟧_p[x:=⟦V⟧_p]} = ⊥, \FLR{⟦N_2⟧_p[x:=⟦V⟧_p]} = L :& ⊥ \\
    \text{else} :& \FLR{⟦N_1⟧_p[x:=⟦V⟧_p]} \FLR{⟦N_2⟧_p[x:=⟦V⟧_p]}
  \end{cases}$ \\
  (Note that we collapsed the $\FLR{⟦N_1⟧_p} = ⊥,\dots$ case.
  We can do that because if $⟦N_1⟧_p = ⊥$ then so does $\FLR{⟦N_1⟧_p[x:=⟦V⟧_p]}$
  and if $⟦N_2⟧_p = L$ then $\FLR{⟦N_2⟧_p[x:=⟦V⟧_p]}$ is also a value.) \\
  By induction, $⟦N_1[x:=V]⟧_p = \FLR{⟦N_1⟧_p[x := ⟦V⟧_p]}$
  and $⟦N_2[x:=V]⟧_p = \FLR{⟦N_2⟧_p[x := ⟦V⟧_p]}$.
\item $y$: trivial because EPP and floor are both no-ops.
\item $(λ y:T_y \DOT N)@\nonempty{p}$:
  \begin{itemize}
  \item If $p \not\in \nonempty{p}$, both sides of the equality are $⊥$.
  \item If $V' = V \mask \nonempty{p}$ is defined, then \\
    $⟦(λ y:T_y \DOT N)@\nonempty{p}[x:=V]⟧_p
    =⟦(λ y:T_y \DOT N[x:=V'])@\nonempty{p}⟧_p
    =  λ y \DOT ⟦N[x:=V']⟧_p$ \\
    and
    $\FLR{⟦(λ y:T_y \DOT N)@\nonempty{p}⟧_p[x := ⟦V⟧_p]}$ \\
    $= \FLR{(λ y \DOT ⟦N⟧_p)[x := ⟦V⟧_p]  }$ \\
    $= \FLR{ λ y \DOT (⟦N⟧_p[x := ⟦V⟧_p]) }$ \\
          $= \FLR{ λ y \DOT (⟦N⟧_p[x := ⟦V'⟧_p])}$ (by Lemma~\ref{theorem:masked}) \\
    $=  λ y \DOT \FLR{(⟦N⟧_p[x := ⟦V'⟧_p])}$ \\
    Then we do induction on $N$ and $V'$.
  \item Otherwise, substitution in the central program is a no-op.  
    \begin{itemize}
    \item $⟦(λ y:T_y \DOT N)@\nonempty{p}[x:=V]⟧_p = ⟦(λ y:T_y \DOT N)@\nonempty{p}⟧_p
      = λ y \DOT ⟦N⟧_p$ \\
      and $\FLR{⟦(λ y:T_y \DOT N)@\nonempty{p}⟧_p[x := ⟦V⟧_p]}
      = \FLR{(λ y \DOT ⟦N⟧_p)[x := ⟦V⟧_p]}
      = \FLR{λ y \DOT (⟦N⟧_p[x := ⟦V⟧_p])}$ \\
      $= λ y \DOT \FLR{⟦N⟧_p[x := ⟦V⟧_p]}$.
    \item Since we already known
      $(λ y:T_y \DOT N)@\nonempty{p}[x:=V] = (λ y:T_y \DOT N)@\nonempty{p}$,
            we can apply Lemma~\ref{theorem:substitution} to $M$ and unpack the typing of
      $M[x:=V] = M$
      to get $\nonempty{p};(y:T_y) ⊢ N : T'$.
  \item By Lemma~\ref{theorem:unused}, we get $N[x:=V] = N$.
    \item By induction on $N$ and $V$, we get
      $\FLR{⟦N⟧_p[x := ⟦V⟧_p]} = ⟦N[x:=V]⟧_p =  ⟦N⟧_p$,
      QED.
    \end{itemize}
  \end{itemize}
\item $\CASE{\nonempty{p}}{N}{x_l}{N_l}{x_r}{N_r}$: (maybe I should work these out more?)
  \begin{itemize}
  \item If $⟦N⟧_p = ⊥$ then $\FLR{⟦N⟧_p[x:=⟦V⟧_p]} = ⊥ = ⟦N[x:=V]⟧_p$ (by induction),
    so both halfs of the equality are $⊥$.
  \item Else if $p \not \in \nonempty{p}$, then we get \\
    $⟦\CASE{\nonempty{p}}{N[x:=V]}{x_l}{N_l'}{x_r}{N_r'}⟧_p
    = \CASE{\nonempty{p}}{⟦N[x:=V]⟧_p}{x_l}{⊥}{x_r}{⊥}$ \\
    and \\
    $\FLR{⟦\CASE{\nonempty{p}}{N}{x_l}{N_l}{x_r}{N_r}⟧_p[x := ⟦V⟧_p]} \\
    = \FLR{(\CASE{\nonempty{p}}{⟦N⟧_p}{x_l}{⊥}{x_r}{⊥})[x := ⟦V⟧_p]} \\
    = \FLR{\CASE{\nonempty{p}}{⟦N⟧_p[x := ⟦V⟧_p]}{x_l}{⊥}{x_r}{⊥}}$. \\
    Since we've assumed $\FLR{⟦N⟧_p[x:=⟦V⟧_p]} \neq ⊥$,
    these are equal by induction.
  \item Else if $V' = V \mask \nonempty{p}$ is defined then we can do induction similar
    similar to how we did for the respective lambda case, except the induction is
    three-way.
  \item Otherwise, it's similar to the respective lambda case, just more verbose.
  \end{itemize}
\item $()@\nonempty{p}$, $\FST{\nonempty{p}}$, $\SND{\nonempty{p}}$,
  $\LOOKUP{i}{\nonempty{p}}$, and $\COMM{s}{\nonempty{r}}$:
  trivial because substitution and floor are no-ops.
\end{itemize}

\begin{lemma}[Weak Completeness]\label{theorem:weak-completeness}
  If $Θ;∅ ⊢ M : T$ and $M \step M'$
  then $⟦M⟧_p \prcstep{μ}{η}^{?} ⟦M'⟧_p$.  
  (\ie it takes zero or one steps to get there.)
\end{lemma}
\subsection{Proof of Lemma~\ref{theorem:weak-completeness}}
If $⟦M⟧_p = ⊥$ then this is follows trivially from Lemma~\ref{theorem:bottom},
so assume it doesn't.
We proceed with induction on the form of $M \step M'$:
\begin{itemize}
\item \textsc{AppAbs}: $M = (λ x:T_x \DOT N)@\nonempty{p} V$,
  and $M' = N[x:=V\mask\nonempty{p}]$.
  By assumption, the lambda doesn't project to $⊥$, so $p \in \nonempty{p}$
  and $⟦M⟧_p \prcstep{∅}{∅} \FLR{⟦N⟧_p[x:=⟦V⟧_p]}$ by \textsc{LAbsApp}. \\
        By Lemma~\ref{theorem:masked} and Lemma~\ref{theorem:distributive-substitution}
  $\FLR{⟦N⟧_p[x:=⟦V⟧_p]} = \FLR{⟦N⟧_p[x:=⟦V\mask\nonempty{p}⟧_p]}
  = ⟦N[x:=V\mask\nonempty{p}]⟧_p = ⟦M'⟧_p$.
\item \textsc{App1}: $M = V N \step V N' = M'$.
  By induction, $⟦N⟧_p \prcstep{μ}{η}^{?} ⟦N'⟧_p$.
  \begin{itemize}
  \item Assume $⟦V⟧_p = ⊥$.
    By our earlier assumption, $⟦N⟧_p \neq ⊥$.
    Since $⟦N⟧_p$ can step; that step justifies a \textsc{LApp1} step
    with the same annotations.
          If $⟦N'⟧_p$ is a value then
    that'll be handled by the floor built into \textsc{LApp1}.
  \item Otherwise, the induction is even simpler,
    we just don't have to worry about possibly collapsing the whole thing to $⊥$.
  \end{itemize}
\item \textsc{App2}:
  $M = N_1 N_2 \step N_1' N_2 = M'$.
  By induction, $⟦N_1⟧_p \prcstep{μ}{η}^{?} ⟦N_1'⟧_p$.
  \begin{itemize}
  \item Assume $⟦N_2⟧_p = L$.
    By our earlier assumption, $⟦N_1⟧_p \neq ⊥$.
    Since $⟦N_1⟧_p$ steps, that step justifies a \textsc{LApp2} step
    with the same annotations.
         If $⟦N_1'⟧_p$ is a value then
    that'll be handled by the floor built into \textsc{LApp2}.
  \item Otherwise, the induction is even simpler.
  \end{itemize}
\item \textsc{Case}: By our assumptions, the guard can't project to $⊥$;
  we just do induction on the guard to satisfy \textsc{LCase}.
\item \textsc{CaseL} (\textsc{CaseR} mirrors):
  $M = \CASE{\nonempty{p}}{\INL V}{x_l}{M_l}{x_r}{M_r}$,
  and $⟦M⟧_p = \CASE{}{\INL ⟦V⟧_p}{x_l}{B_l}{x_r}{B_r}$.
  $⟦M⟧_p \prcstep{∅}{∅} \FLR{B_l[x_l := ⟦V⟧_p]}$ by \textsc{LCaseL}.
  $M' = M_l[x_l := V\mask\nonempty{p}]$.
  If $p \in \nonempty{p}$
  then $B_l = ⟦M_l⟧_p$
        and by Lemma~\ref{theorem:masked} and Lemma~\ref{theorem:distributive-substitution}
  $\FLR{B_l[x_l := ⟦V⟧_p]} = \FLR{⟦M_l⟧_p[x_l := ⟦V⟧_p]}
  = \FLR{⟦M_l⟧_p[x_l := ⟦V\mask\nonempty{p}⟧_p]}
  = ⟦M_l[x_l := V\mask\nonempty{p}]⟧_p
  = ⟦M'⟧_p$. \\
  Otherwise, $B_l[x_l := ⟦V⟧_p] = ⊥$
        and by \textsc{TCase}, Lemma~\ref{theorem:substitution}, and Lemma~\ref{theorem:cruft},
  $⟦M'⟧_p = ⊥$.
\item \textsc{Proj1}: $M = \FST{\nonempty{p}} (\PAIR V_1 V_2)$
  and $M' = V_1 \mask \nonempty{p}$.
  Since we assumed $⟦M⟧_p \neq ⊥$, $p \in \nonempty{p}$. \\
  $⟦M⟧_p = \FST{} \FLR{\PAIR ⟦V_1⟧_p ⟦V_2⟧_p} = \FST{} (\PAIR ⟦V_1⟧_p ⟦V_2⟧_p)$
        by Lemma~\ref{theorem:existence} and \textsc{TPair}.
  This steps by \textsc{LProj1} to $⟦V_1⟧_p$,
        which equals $⟦M'⟧_p$ by Lemma~\ref{theorem:masked}.
\item \textsc{Proj2}, \textsc{ProjN}: Same as \textsc{Proj1}.
\item \textsc{Com1}: $M = \COMM{s}{\nonempty{r}} ()@\nonempty{p}$
  and $M' = ()@\nonempty{r}$.
  \begin{itemize}
  \item $s = p$ and $p \in \nonempty{r}$:
    By \textsc{MVUnit}, $p \in \nonempty{p}$,
    so $⟦M⟧_p = \SEND{\nonempty{r} ∖ \set{p}}^{\ast} ()$,
    which steps by \textsc{LSendSelf} (using \textsc{LSend1}) to $()$.
    $⟦M'⟧_p = ()$.
  \item $s = p$ and $p \not\in \nonempty{r}$:
    By \textsc{MVUnit}, $p \in \nonempty{p}$,
    so $⟦M⟧_p = \SEND{\nonempty{r}} ()$,
    which steps by \textsc{LSend1} to $⊥$.
    $⟦M'⟧_p = ⊥$.
  \item $s \neq p$ and $p \in \nonempty{r}$:
    $⟦M⟧_p = \RECV{s} ⟦()@\nonempty{p}⟧_p$,
    which can step
    (arbitrarily, but with respective annotation)
    by \textsc{LRecv} to $⟦M'⟧_p$.
  \item Otherwise, we violate our earlier assumption.
  \end{itemize}
\item \textsc{ComPair}, \textsc{ComInl}, and \textsc{ComInr}:
  Each uses the same structure of proof as \text{Com1},
  using induction between the cases
  to support the respective process-semantics step.
\end{itemize}

\subsection{Proof of Theorem~\ref{theorem:completeness}}
By case analysis on the semantic step $M \step M'$:
\begin{itemize}
\item \textsc{AppAbs},
  \textsc{CaseL},
  \textsc{CaseR},
  \textsc{Proj1},
  \textsc{Proj2},
  and \textsc{ProjN}:
  Necessarily, the set of parties $\nonempty{p}$ for whom
  $⟦M⟧_{p\in\nonempty{p}} \neq ⊥$ is not empty.
  For every $p \in \nonempty{p}$,
        by Lemma~\ref{theorem:weak-completeness} $⟦M⟧_p \prcstep{∅}{∅}^{?} ⟦M'⟧_p$
  (checking the cases to see that the annotations are really empty!).
  By \textsc{NPro}, each of those is also a
  network step,
        which by Lemma~\ref{theorem:parallelism} can be composed in any order to get
  $⟦M⟧ \netstep{}{∅}^{\ast} \mathcal{N}$.
  For every $p \in \nonempty{p}$,
  $\mathcal{N}(p) = ⟦M'⟧_p$,
        and (by Lemma~\ref{theorem:bottom}) for every $q \not\in \nonempty{p}$,
  $\mathcal{N}(q) = ⊥ = ⟦M'⟧_q$,
  Q.E.D.
\item \textsc{Com1},
  \textsc{ComPair},
  \textsc{ComInl},
  and \textsc{ComInr}:
  $M = \COMM{s}{\nonempty{r}} V$.
  By the recursive structure of \textsc{Com1}, \textsc{ComPair}, \textsc{ComInl},
  and \textsc{ComInr}, $M'$ is some structure of
  $\set{\PAIR, \INL{}, \INR{}, ()@\nonempty{r}}$,
  and $⟦M'⟧_{r\in\nonempty{r}} = ⟦V⟧_s$.
  For every $q \not\in \nonempty{r} ∪ \set{s}$, $⟦M⟧_q = ⊥ = ⟦M'⟧_q$
        by Lemma~\ref{theorem:bottom}.
  Consider two cases:
  \begin{itemize}
  \item $s \not\in \nonempty{r}$: \\
      By Lemma~\ref{theorem:weak-completeness}
    $⟦M⟧_s = \SEND{\nonempty{r}} ⟦V⟧_s
    \prcstep{\set{(r, ⟦V⟧_s) \mid r \in \nonempty{r}}}{∅} ⊥$.\\
    By the previously mentioned structure of $M'$, $⟦M'⟧_s = ⊥$. \\
    For every $r \in \nonempty{r}$,
    by Lemma~\ref{theorem:weak-completeness}
    $⟦M⟧_r = \RECV{s} ⟦V⟧_r
    \prcstep{∅}{\set{(s,⟦V⟧_s)}} ⟦V⟧_s = ⟦M'⟧_{r}$. \\
    By \textsc{NPro},
    $s[⟦M⟧_s] \netstep{s}{\set{(r, ⟦V⟧_s) \mid r \in \nonempty{r}}} s[⊥=⟦M'⟧_s]$.\\
    This composes in parallel with each of the $r_{\in\nonempty{r}}[⟦M⟧_r]$
    by \textsc{NCom} in any order until the unmactched send is empty.
    Everyone in and not-in $\nonempty{r} ∪ \set{s}$ has stepped, if needed,
    to the respective projection of $M'$.
  \item $s \in \nonempty{r}$: Let $\nonempty{r_0} = \nonempty{r} ∖ \set{s}$. \\
    By Lemma~\ref{theorem:weak-completeness}
    $⟦M⟧_s = \SEND{\nonempty{r_0}}^{\ast} ⟦V⟧_s
    \prcstep{\set{(r, ⟦V⟧_s) \mid r \in \nonempty{r_0}}}{∅} ⟦V⟧_s
    = ⟦M'⟧_{s\in \nonempty{r}}$. \\
    For every $r \in \nonempty{r_0}$,
    by Lemma~\ref{theorem:weak-completeness}
    $⟦M⟧_r = \RECV{s} ⟦V⟧_r
    \prcstep{∅}{\set{(s,⟦V⟧_s)}} ⟦V⟧_s = ⟦M'⟧_{r}$. \\
    We proceed as in the previous case.
  \end{itemize}
\item \textsc{App1} (\textsc{App2} and \textsc{Case} are similar):
  $M = V N$.
  By induction, $⟦N⟧ \netstep{}{∅}^{\ast} ⟦N'⟧$.
  Every $N$ step in that process in which a single party advances by \textsc{NPro}
  can justify a corresponding $M$ step by \textsc{LApp1}.
  \textsc{NCom} steps are basically the same: each of the participating parties will
  justify a \textsc{LApp1} $M$ step with a $N$ step;
  since this doesn't change the send \& receive annotations,
  the cancellation will still work.
\end{itemize}

\section{A more complex translation from Chorλ}\label{sec:translation}
Figure~\ref{fig:chor-lambda-complex} shows a Chorλ choreography that actually leverages the communication efficiency of the select-\&-merge paradigm,
and which is deliberately obnoxious in its asymmetric flow.
Figure~\ref{fig:our-complex} is a human translation of that same choreography into \FuncLang.
It's verbose because it closely follows the strategy described in Section~\ref{sec:chor-lambda}; a fully mechanized translation would be even more verbose.

\begin{figure}[tbhp]
    \begin{mdframed}
    \begin{minted}[xleftmargin=10pt,linenos]{bash}
let m1 = com[q][p] n_q1;
let (cache1, flag1) = case ( first_secret[p] ()@[p] ) of
  Inl _ => let (c1_, f1_) = case ( second_secret[p] ()@[p] ) of
             Inl _ => let w = m1;
                      (Inl w, Inl ()@[p]);
             Inr _ => let w = m1;
                      let y = 2@p;
                      (Inr (Pair w y), Inl ()@[p]);
           (Inl c1_, f1_);
  Inr _ => let (c1_, f1_) = let w = m1;
                            case ( second_secret[p] ()@[p] ) of
                              Inl s => (Inl (Pair w s), Inl ()@[p]);
                              Inr _ => (Inr w, Inr ()@[p]);
           (Inr c1_, f1_);
let f1_ = com[p][p,q] flag1;
case f1_ of Inl _ => let (cache2, m2) = case cache1 of
                        Inl c1l => let (c2_, m2_) = case c1l of
                                     Inl c1ll => let w = c1ll;
                                                 (Inl w, w + 1@[p]);
                                     Inr c1lr => let (Pair w y) = c1lr;
                                                 (Inr (Pair w y), w + y);
                                   (Inl c2_, m2_);
                        Inr c1r => let (c2_, m2_) = case c1r of
                                     Inl c1rl => let (Pair w s) = c1rl;
                                                 (Pair w s, 5@[p]);
                                     Inr c1rr => (DEFAULT, DEFAULT); # DEAD BRANCH
                                   (Inr c2_, m2_);
                     let _ = com[p][q] m2;
                     case cache2 of
                       Inl c2l => case c2l of
                         Inl c2ll => let w = c2ll;
                                     w + 1@[p];
                         Inr c2lr => let (Pair w y) = c2lr;
                                     w;
                       Inr c2r => let (Pair w s) = c2r;
                                  s;
            Inr _ =>  let cache2 = case cache1 of
                        Inl c1l => DEFAULT; # DEAD BRANCH
                        Inr c1r => case c1r of
                          Inl c1rl => DEFAULT; # DEAD BRANCH
                          Inr c1rr => let w = c1rr;
                                      w;
                      let m2 = com[q][p] n_q2;
                      let w = cache2;
                      let z = m2;
                      w + z
    \end{minted}
        \caption{An algorithmic \FuncLang translation of the choreography from Figure~\ref{fig:chor-lambda-complex}.}
    \label{fig:our-complex}
    \Description[46 lines of He-Lambda-Small code. The program flow from the earlier program is still present, but hard to see.]
                {46 lines of He-Lambda-Small code.
                The program flow from the earlier program is still present,
                but hard to see because of all the temporary variables as execution jumps in and out of sequential and nested case expressions.}
    \end{mdframed}
\end{figure}

\end{document}